\documentclass[11pt,a4paper]{article}
\usepackage{jcappub}
\usepackage[sort&compress,numbers]{natbib}
\usepackage[T1]{fontenc}

\usepackage[utf8]{inputenc}
\usepackage{textcomp}
\usepackage{gensymb}
\usepackage{amsmath}
\usepackage{amssymb}
\usepackage{xspace}
\usepackage{nicefrac}
\usepackage{xcolor}
\usepackage{graphicx}
\usepackage{soul}
\usepackage{hyperref}
\usepackage[caption=false]{subfig}
\usepackage{multirow}
\usepackage{cleveref}
\usepackage[titletoc,toc,page]{appendix}

\title{Nearby active galactic nuclei and starburst galaxies as sources of the measured UHECRs anisotropy signal}

\date{\today}
\author{Cainã de Oliveira}
\emailAdd{caina.oliveira@usp.br}
\author{and Vitor de Souza}
\emailAdd{vitor@ifsc.usp.br}
\affiliation{Instituto de F\'isica de S\~ao Carlos, Universidade de S\~ao Paulo, Av. Trabalhador S\~ao-carlense 400, S\~ao Carlos, Brasil.}

\abstract{
The Pierre Auger and the Telescope Array observatories have measured independent and statistical significant anisotropy in the arrival direction of ultra-high-energy cosmic rays (UHECR). Three hotspot regions with relative excess of events and a dipole signal have been identified in different regions of the sky and energy ranges. In this paper, we investigate the conditions under which these anisotropy signal could be generated by nearby ($<$23~Mpc) active galactic nuclei (AGN) and/or starburst galaxies (SBG). We studied a wide range of possibilities including injected nuclei (p, He, N, Si, and Fe), three UHECR luminosity proxies and three extragalactic magnetic field models. The results shows that both local AGN and SBG are needed to describe all the anisotropy signal. The contribution of AGN to hotspots and to the generation of the dipole is dominant in most cases. SBG is required only to explain the hotspot measured by the Telescope Array Observatory.
}
\begin{document}
%\linenumbers
\maketitle
\flushbottom

\section{Introduction}

%================
%Abertura e Guia da introdução

The existence of ultra-high-energy cosmic rays (UHECR, with energy above $10^{18}$ eV = 1 EeV) has been known since the 1960s~\citep{PhysRevLett.10.146}. Even after decades of theoretical efforts and several dedicated experiments~\citep{RevModPhys.72.689}, the sources and acceleration mechanisms of these particles remain unknown~\citep{alves2019open}. The difficulty in identifying the sources is related to the large uncertainties in the data, the small number of events detected so far, and by the lack of knowledge about the cosmic magnetic fields.

%========================
%Teoria de AGN e SBR

The number of potential sources of UHECR is strongly restricted by the Hillas condition~\citep{hillas_1984}. Among the most probably candidates are active galactic nuclei (AGNs) and starburst galaxies (SBGs)~\citep{PhysRevD.99.063012,alves2019open}. The particle acceleration in AGNs has been widely studied by many authors~\citep{ginzburg1963cosmic,Caprioli_2015,liu2017particle,PhysRevD.97.023026,eichmann2018ultra,matthews2019ultrahigh,eichmann2019high}. The presence of relativistic jets emanating from supermassive black holes creates extreme environments in which the particle's acceleration occurs~\citep{dermer2009ultra,matthews2019ultrahigh}. The radio galaxies, Centaurus~A (Cen~A, NGC~5128), Virgo~A (Vir~A, M87), and Fornax~A (For~A, NGC~1316) are the closest and most powerful nearby AGNs~\citep{van2012radio,cavagnolo2010}.

SBGs drive nuclear outflows in the form of powerful and magnetized winds that constitute a potential site for particle acceleration~\citep{PhysRevD.60.103001, PhysRevD.97.063010,PhysRevD.99.063012}. The details about the efficiency of the superwind in accelerating particles at the highest energies is still under debate~\citep{romero2018_sbgwinds,matthews2018fornax,PhysRevD.77.123003,10.1093/mnras/stac031}. Besides that, SBGs have adequate environments for frequent extreme events in which particle acceleration is expected, such as gamma-ray bursts, trans-relativistic supernovae, and hypernovae~\citep{PhysRevD.100.103004,PhysRevD.97.083010,arxiv.1610.00944}.

%===========================
%Dados

Recently, the Pierre Auger~\citep{bib:auger} and the Telescope Array~\citep{bib:ta} Observatories have improved the UHECR data quality and have increased the number of detected events allowing precision studies. The Pierre Auger Observatory detected a large-scale dipolar modulation in the arrival direction of UHECR with energies above 4, 8, and 32~EeV~\citep{1266,Aab_2018}. If events with energy above 4~EeV are analyzed, the dipole amplitude increases with energy and the statistical significance of the signal has increased with the accumulation of more data~\citep{Aab_2018, Aab_2020_raAnisotropies, deAlmeida:20212Z}. The interpretations of these data favor extragalactic sources of UHECR~\citep{1266}. A similar study done by the Telescope Array Collaboration reports a dipole structure compatible with isotropy and with the Auger dipole at the same time~\citep{Abbasi_2020_largescale_anisotropy}.

In addition to the large-scale anisotropy, the Pierre Auger and the Telescope Array Observatories detected regions in the sky with a relative excess of events. The Pierre Auger Collaboration reported two hotspots regions (HS1 and HS2) for events with energy above 60~EeV~\citep{aab2018indication} centered approximately at (305$^\circ$,25$^\circ$) and (290$^\circ$,-70$^\circ$)~\citep{de_Oliveira_2022} in galactic coordinates. The Telescope Array Collaboration reported one hotspot (HS3) for events with energy above 57~EeV centered at (146$^\circ$.7,43$^\circ$.2) in equatorial coordinates~\citep{Abbasi_2014}. Cen~A has been considered the main candidate to explain one of the hotspot~\citep{ginzburg1963cosmic,ROMERO1996279,biermann2012centaurus,liu2012excess,wykes2013mass,Farrar_2013,wykes2018uhecr,deoliveira2021probing,de_Oliveira_2022,caccianiga2019anisotropies}. The absence of hotspots in the Vir~A direction can be explained by the effect of the extragalactic magnetic fields in the propagation of the particles~\citep{Dolag_2009,de_Oliveira_2022}. It has been suggested that For~A can increase the correlation of the Auger data with $\gamma$-ray AGN catalogs~\citep{matthews2018fornax,de_Oliveira_2022}.

The Pierre Auger Observatory and the Telescope Array data have also been compared to source catalogs.
The Pierre Auger data shows a correlation with $\gamma$-ray AGN catalogs (2.7$\sigma$) and with SBGs catalogs (4.0$\sigma$) against the isotropic hypothesis~\citep{aab2018indication}. The Telescope Array data is not able to discriminate between the SBG catalog and an isotropic sky~\citep{Abbasi_2018}.

%==========================
%Campo magnético

Extragalactic and galactic magnetic fields (EGMF and GMF) deflect UHECR during their propagation. The lack of information about these cosmic magnetic fields represents one major barrier to determining the source~\citep{ERDMANN201654,Farrar_2019,https://doi.org/10.48550/arxiv.astro-ph/0309695,PhysRevD.68.043002,Dolag_2005,Dolag_2009,PhysRevD.96.023010,de_Oliveira_2022}. Some information about the intensity of the EGMF be obtained by experimental techniques and little is known about its structure~\citep{Subramanian_2016}. Nowadays, computational simulations have been used to infer more properties of the EGMF~\citep{PhysRevD.68.043002,Dolag_2005,hackstein2018simulations}. The data about the GMF are less scarce than of the EGMF and observationally driven models have been developed~\citep{sun_2008_gmf,Jansson_2012a,Jansson_2012b,Pshirkov_2011_gmf}. However, significant improvements are needed for a better description of the UHECR deflections in the GMF~\citep{ERDMANN201654,Farrar_2019}.

%===========================
%Fontes próximas

UHECRs interact with background photons resulting in energy loss via $e^+e^-$ pair production and pion photoproduction. These processes are energy-dependent and limit the maximal distance of the sources, indicating the need for nearby sources of UHECR~\citep{bib:zk,Greisen1966, DELIGNY2004609,taylor2011need,AlvesBatista:2014tdg, lang2020revisiting}. The increase of the magnitude of the dipole modulation with energy also corroborates a growing contribution of nearby sources to the flux of UHECR~\citep{lang2020_dipole}. The necessity of nearby sources has increased the expectation for an UHECR astronomy~\citep{RevModPhys.71.S165}.

%=========================
%Conclusão da introdução: falar do seu trabalho.

Recently, we have reported~\cite{de_Oliveira_2022} that the Pierre Auger dipole above 32 EeV can be understood based on an excess of events coming from the three closest AGNs (Cen~A, Vir~A, and For~A). In this paper we extend the analysis, examining the effects of the population of the nearest SBGs and of a combined approach of SBGs and AGNs. For the first time, we report that a scenario based alone on SBGs is unlikely to reproduce the dipole direction measured by the Pierre Auger Observatory, providing another layer of evidence for the particle's acceleration in radio galaxies. In section~\ref{sec:method} we present the details of source selection and computational simulation. We also discuss the energy spectrum imposed on each source. In section~\ref{sec:arrival} general remarks about the angular contribution for the flux of UHECR coming from each source are discussed, including the possibility of contribution to each measured hotspot. The dipole generation in each scenario is discussed in section~\ref{sec:dipoles}, in which we compare the dipole signal generated by SBGs only, AGNs only, and the combination of SBGs and AGNs. Finally, we conclude in section~\ref{sec:conclusion}.

%%See
%https://iopscience.iop.org/article/10.1088/1742-6596/2156/1/012007/pdf
%https://iopscience.iop.org/article/10.1088/1742-6596/1468/1/012078/pdf
%https://indico.ific.uv.es/event/6178/contributions/15762/attachments/9382/12308/TA_TALE_TAx4_Latest_Resuts.pdf

\section{Method}
\label{sec:method}

\subsection{Source selection and UHECR injection}
In this work, we consider nearby AGNs and SBGs as sources of UHECR and explore the arrival direction signals generated in different EGMF and compositions. As in \citet{de_Oliveira_2022} Cen~A, Vir~A, and For~A were considered the main AGNs contributing to the anisotropic signal in the arrival direction of UHECR~\citep{ginzburg1963cosmic,matthews2018fornax,eichmann2018ultra,eichmann2019high}. The nineteen SBGs closer than 23~Mpc in the catalog used in~\citet{aab2018indication} were selected for the analysis in this work.

The sources are supposed to emit a power-law spectrum with a charge-dependent exponential cutoff~\citep{supanitsky2013upper}. The standard spectral index $-2$ provided by the first-order Fermi acceleration in strong shocks was used~\citep{kotera2011astrophysics}. A cutoff rigidity of 50~EV was selected based on the suppression energy of the UHECR spectrum measured by the Pierre Auger Observatory~\citep{PhysRevD.102.062005}.

The radio~\citep{van2012radio} and $\gamma$-ray~\citep{Ackermann_2012,sahakyan2018} luminosity of each galaxy were used as proxies for the UHECR luminosity ($L_{CR}$). The radio luminosity data at 1.4 GHz ($L_{radio}$) was directly taken from~\citet{van2012radio}. The $\gamma$-ray luminosity between 0.1 and 100 GeV ($L_{\gamma}$) was taken from~\citep{Ackermann_2012,sahakyan2018}. The $\gamma$-ray luminosity for SBGs was corrected by the updated distance values found in~\citet{aab2018indication}. The value of $L_{\gamma}$ for For~A was estimated using the photon flux and spectral index between 1 and 100 GeV published by the Fermi-LAT Collaboration~\footnote{\url{https://www.ssdc.asi.it/fermi3lac/}}. Table~\ref{tab:sources} summarizes the main properties of the SBGs and AGNs used in the analysis.

\subsection{Extragalactic propagation}
The UHECR propagation in the extragalactic environment was simulated using the CRPropa3 framework~\citep{batista2016crpropa}. From each source were injected $10^8$ events of each proton (p), He, N, Si, and Fe nuclei with energies between 8 and 1000~EeV. In the simulations, the events are isotropically emitted by the sources with a distribution following a power-law in energy with spectral index $-1$, to guarantee equal statistical fluctuations at all energies. After the detection, each particle receives an energy-dependent weight to generated the spectrum of interest~\citep{eichmann2018ultra,de_Oliveira_2022}.

During the extragalactic propagation, it is necessary to take into account the interactions with background photons and the deflections due to the EGMF. The simulations include the presence of the cosmic microwave background and the extragalactic background light model of \citet{gilmore2012semi}. The UHECR interactions with the photon backgrounds include $e^+e^-$ pair production, pion photoproduction, and photodisintegration. Nuclear decay and adiabatic losses are also considered.

The extragalactic magnetic field models developed by \citet{hackstein2018simulations} were employed in the simulations. We select the same models used in \citet{de_Oliveira_2022}: AstrophysicalR (AstroR), Primordial (Prim), and Primordial2R (Prim2R). These models cover a wide range of field intensity and level of structure of the EGMF.

The particles were followed until they left a box of size twice the source distance or reached the observer sphere, located in the Milky Way. To maximize the number of detected particles and keep the arrival uncertainties smaller than $1^\circ$, the radius of the detector sphere ($r_{obs}$) was chosen based on the distance ($D$) of the source ($r_{obs} = D \sin{(1^\circ)}$). To ensure that a representative number of events will be detected, the minimum value of $r_{obs}$ was taken as 100~kpc. The effect of differences in the observer area on the detected flux to each source was corrected by applying a $r_{obs}^{-2}$ weight. A discussion about the size of the observer can be found in \citet{de_Oliveira_2022}.

Due to its small size ($\sim$30~kpc) compared to the distance to extragalactic sources ($\sim$Mpc), the energy losses can be ignored in the interior of the Milky Way. The effect of the GMF on the arrival directions of UHECR cannot be ignored. It was accounted using a parametrization of the \citet{Jansson_2012a} GMF model (JF12)~\citep{BRETZ2014110}. Both the regular and random components of JF12 were considered, using the module GalacticLens of CRPropa 3.

%========================================================
\section{Results I: Hotspots generated by AGN\lowercase{s} and/or SBG\lowercase{s}}
\label{sec:arrival}

\begin{figure}
  \centering
  \includegraphics[width=1.0\columnwidth]{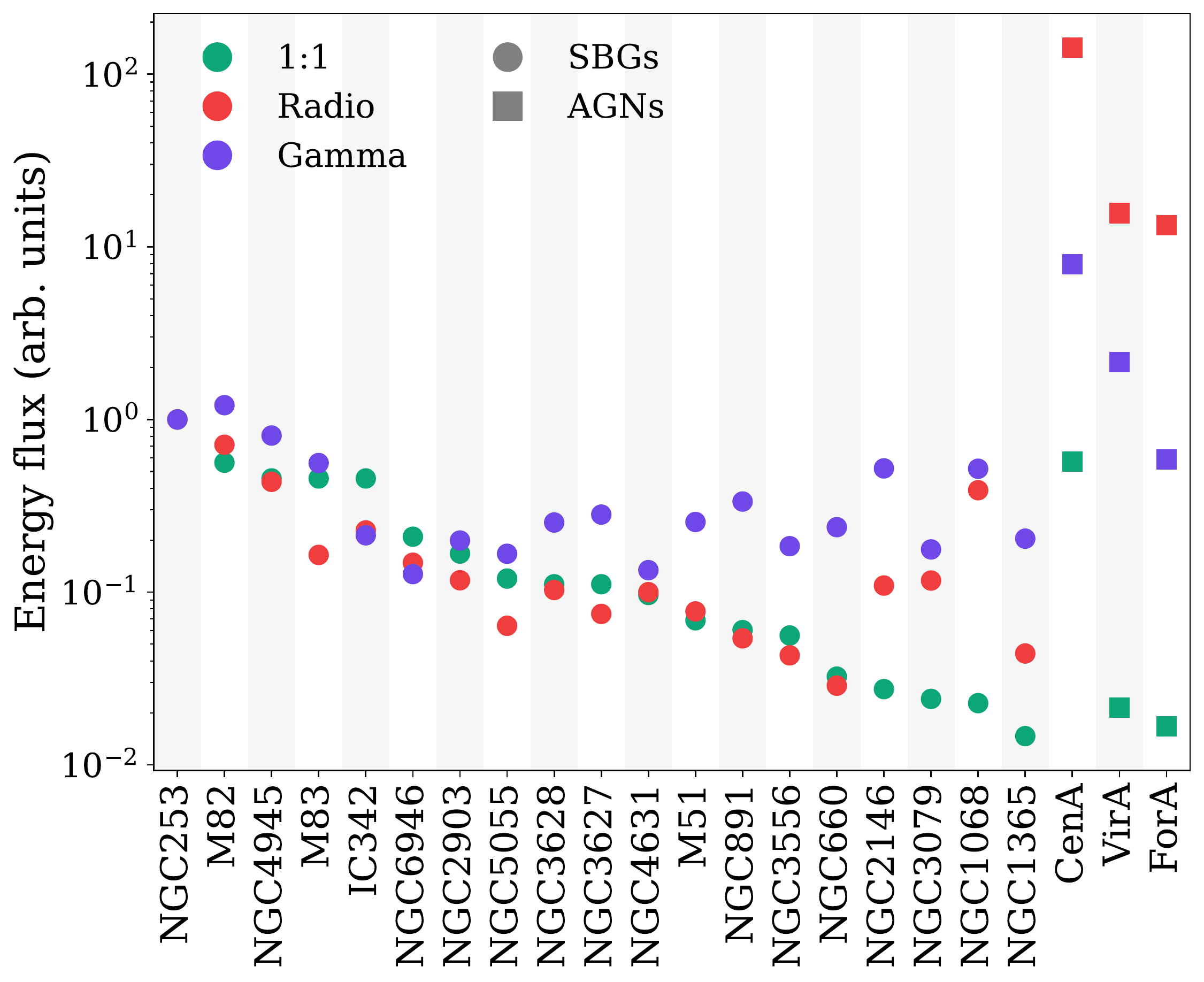}
  \caption{Energy flux at Earth from each source evaluated as $L_{CR}/D^2$. Sources are plotted in order of distance from Earth inside its classification SBGs (circles) and AGNs (squares). NGC~253, the closest sources, has its flux arbitrarily set to one. Three proxies to the UHECR luminosity are presented (1:1, Radio, and Gamma).}
  \label{fig:Eflux_models}
\end{figure}

Figure \ref{fig:Eflux_models} shows the relative contribution of each source evaluated as $L_{CR}/D^2$. NGC~253, the closest source, has its flux arbitrarily set to one. Three normalizations for the flux of energy leaving the sources are used: a) equal flux (1:1), b) radio flux, and c) $\gamma$-ray flux as explained above. The use of electromagnetic luminosities as gauges to $L_{CR}$ changes considerably the dominant sources. In the case that $L_{CR}$ scales with $L_{radio}$, the radio galaxies strongly dominate over the SBGs. If the $L_{CR}$ scales with $L_{\gamma}$, Cen~A remains the main source followed by VirA and the closest SBGs (M82 and NGC~253). Distant SBGs sources (16.3 Mpc) can have a contribution to the energy flux compared to the nearby sources if $L_{radio}$ or $L_{\gamma}$ are used as a proxy (e.g. \cite{LAMASTRA201916}).

\begin{figure}
  \centering
  \includegraphics[width=1.0\columnwidth]{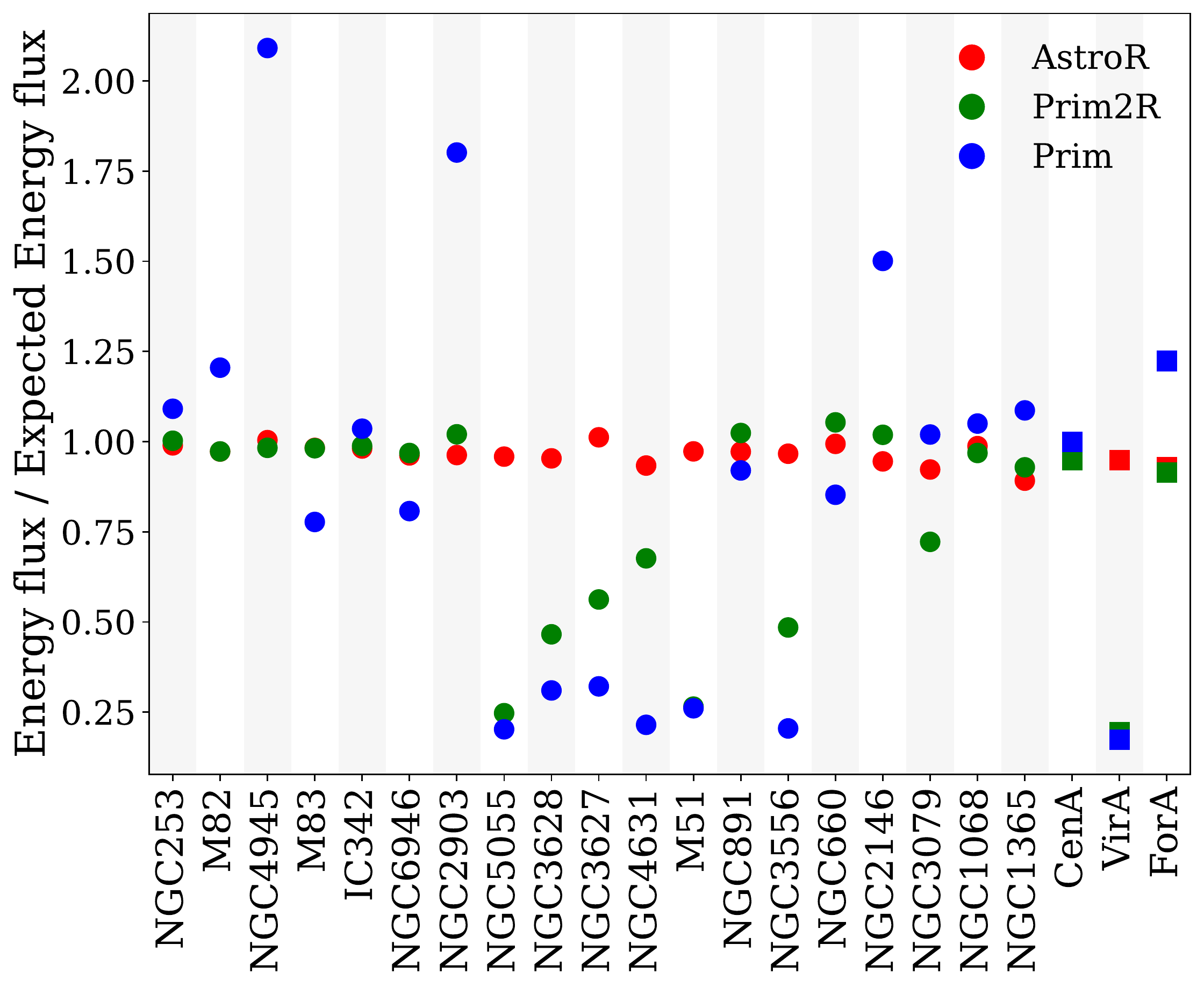}
  \caption{Ratio between the energy flux detected at Earth and the expected energy flux from the sources. Only protons emitted by the sources are considered. The EGMFs models AstroR (red), Prim2R (green), and Prim (blue) are showed. SBGs (circles) and AGNs (squares) are showed.}
  \label{fig:magnetic_effects}
\end{figure}

The influence of the EGMF model is illustrated in figure~\ref{fig:magnetic_effects}. NGC~253, the closest source, has its value arbitrarily set to one when the AstroR EGMF is used. The AstroR EGMF model has a very small ($<10$\%) effect on the energy flux received on Earth as emitted by any of the sources. In this case, the closer sources have a more important contribution, being dominated by NGC~253 among the SBGs, and by Cen~A among the AGNs. The Prim2R EGMF model suppresses the energy flux of NGC~5055, NGC~3628, NGC~3627, NGC~4631, M51, NGC~3556, NGC~3079, and Vir~A. In the Prim EGMF model, the magnetic deflections can suppress or enhance the contribution from a source. The suppression occurs to the same sources affected by the Prim2R, in addition to M83, NGC~6946, and NGC~660. The enhancement happens to M82, NGC~4945, NGC~2903, NGC~2146, and For~A.

Figures \ref{fig:arrival:60EeV:AGN} and \ref{fig:arrival:60EeV:SBG} shows the arrival directions map for events with energies above 60~EeV when local AGNs and SBGs were considered the sources of UHECR, respectively. The color scale indicates the number of events on a logarithm scale arriving at Earth. In each line, one EGMF model is presented. In each column, the injected nuclei are shown: proton, nitrogen, and iron. The position of the sources is shown only in the first diagram for sake of clarity. The hotspot regions measured by the Pierre Auger (HS1 and HS2) and the Telescope Array (HS3) are also shown. The majority of suppressed sources are located in regions of large ($>60^\circ$) galactic latitudes. The effect of the GMF is discussed in the appendix~\ref{sec:appendix}.

\begin{figure}
  \centering
  \includegraphics[width=1.0\columnwidth]{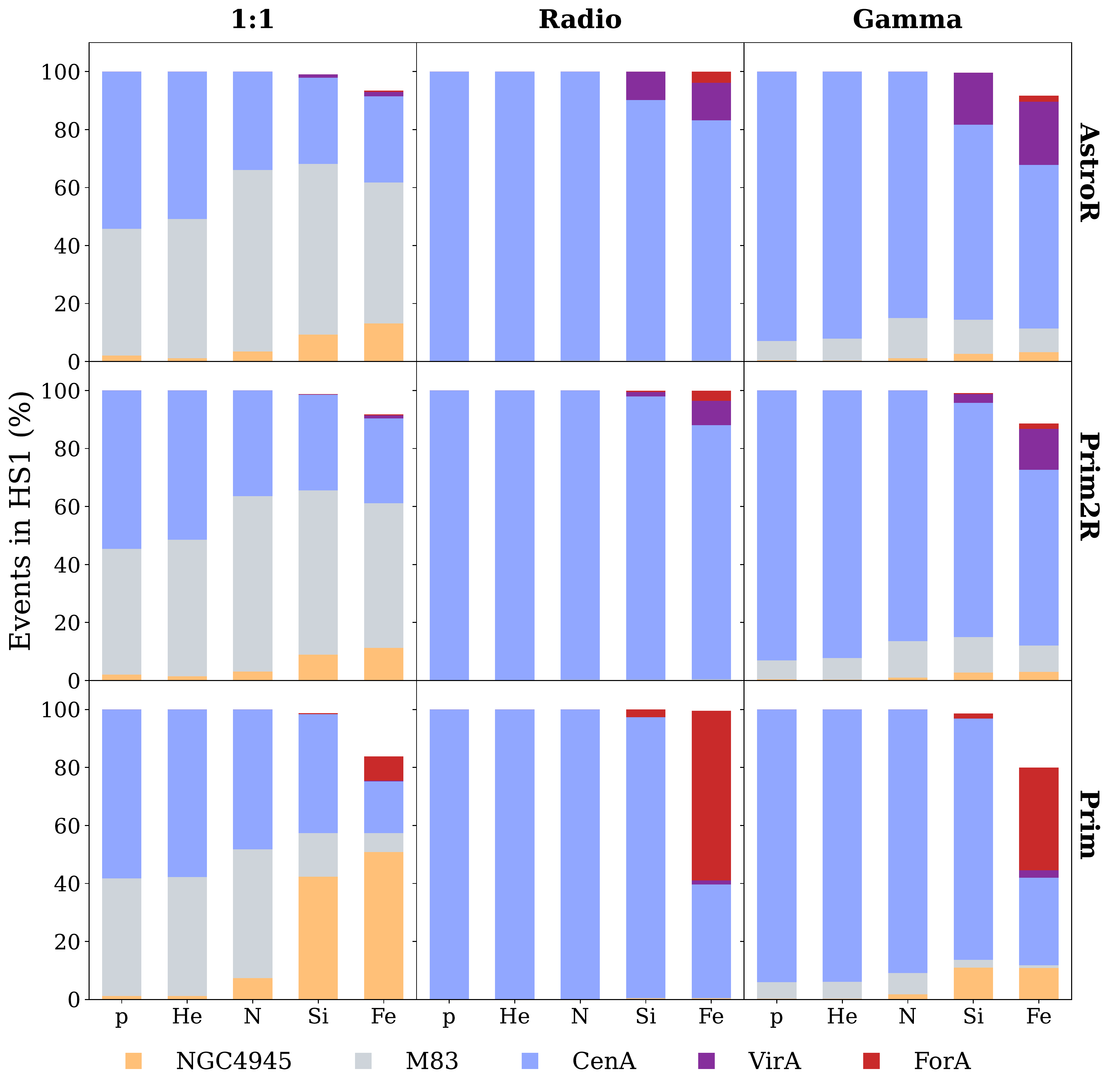}
  \caption{Contribution of sources to the HS1 separed by each UHECR luminosity proxy (1:1, Radio, and Gamma) and EGMF (AstroR, Prim2R, and Prim) models. Only sources contributing with more than 10\% are showed.}
  \label{fig:contribution_HS1}
\end{figure}

HS1 is populated mainly by M83 and Cen~A. When light nuclei are injected (p, He, N), M83 and Cen~A generate more than 95\% of the events populating the HS1 as shown in figure~\ref{fig:contribution_HS1}. When heavier nuclei are injected (Si and Fe), NGC~4945, For~A, and Vir~A also contribute to HS1, but the exact contribution of each source becomes highly dependent on the EGMF model.

\begin{figure}
  \centering
  \includegraphics[width=1.0\columnwidth]{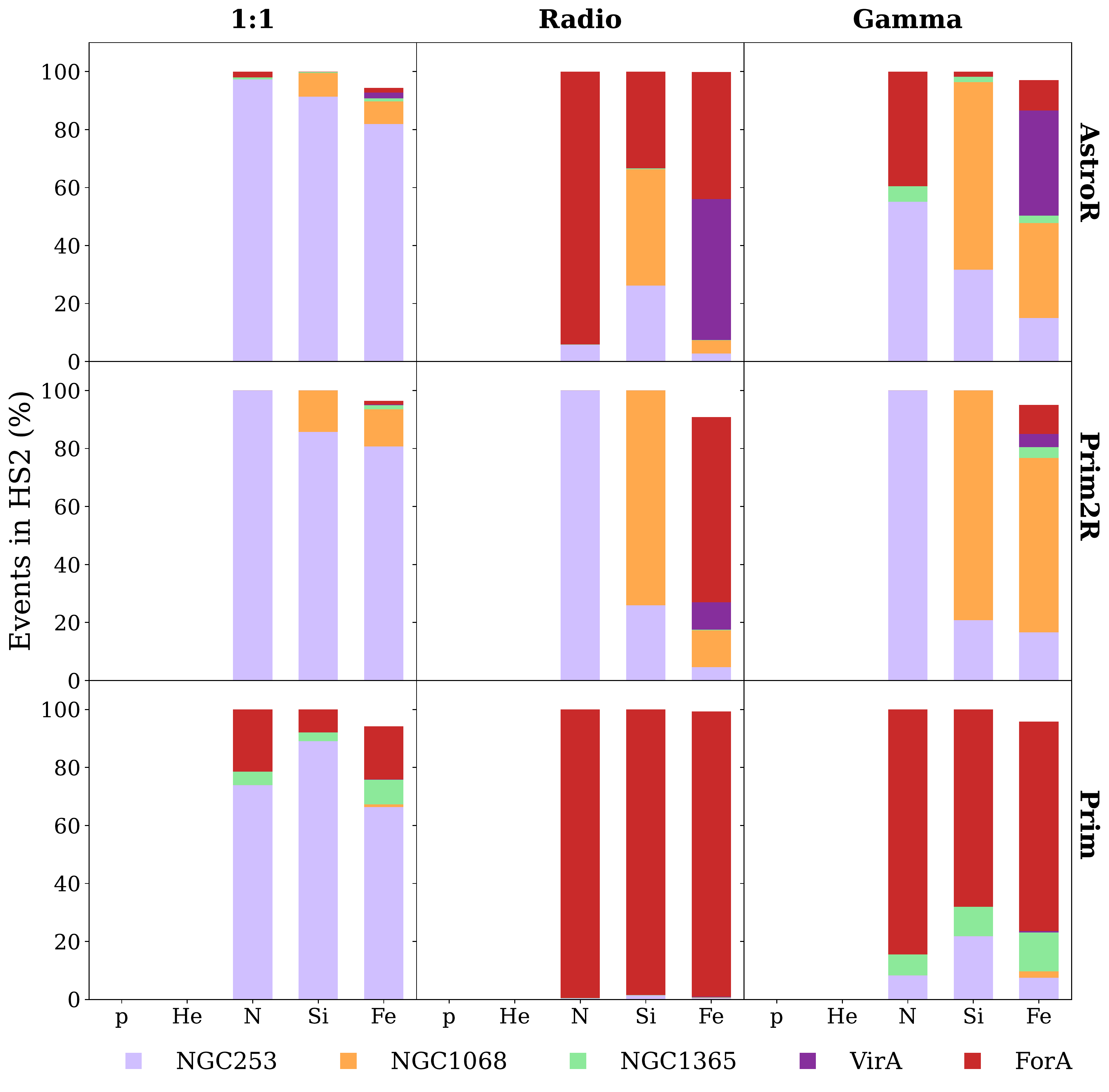}
  \caption{Contribution of sources to the HS2 separed by each UHECR luminosity (1:1, Radio, and Gamma) and EGMF (AstroR, Prim2R, and Prim) models. Only sources contributing with more than 10\% are showed.}
  \label{fig:contribution_HS2}
\end{figure}

As shown in figure \ref{fig:contribution_HS2}, the HS2 is populated primarily by NGC~253, NGC~1068, and For~A. The dominant source is highly dependent on the EGMF and composition injected. Note that light nuclei (p and He) injected by these sources do not reach the HS2 location.

\begin{figure}
  \centering
  \includegraphics[width=1.0\columnwidth]{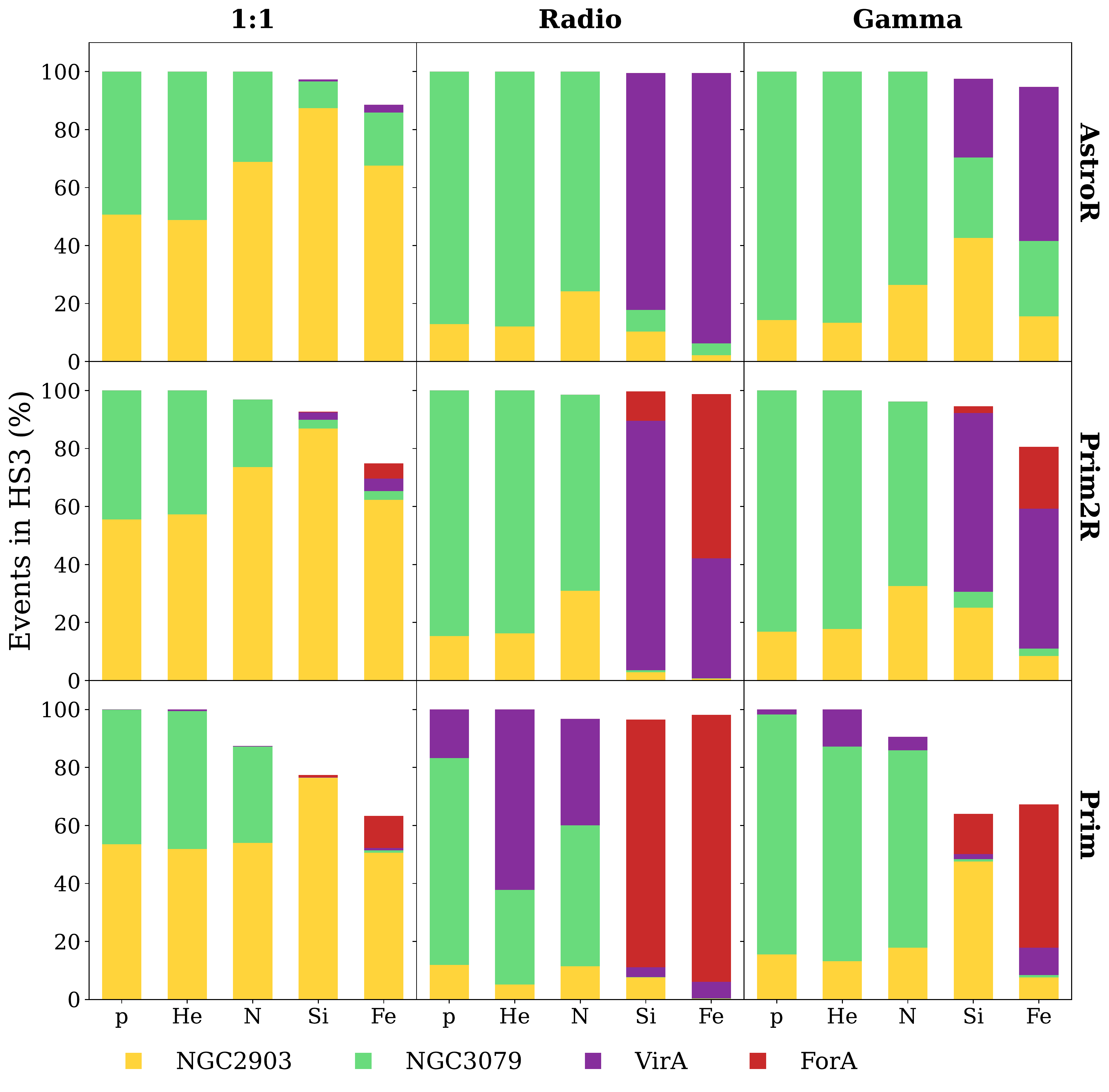}
  \caption{Contribution of sources to the HS3 separed by each UHECR luminosity (1:1, Radio, and Gamma) and EGMF (AstroR, Prim2R, and Prim) models. Only sources contributing with more than 10\% are showed.}
  \label{fig:contribution_HS3}
\end{figure}

HS3 is mostly populated by NGC~2903 and NGC~3079 independently of the EGMF model in the case of light and intermediate composition, as shown in figure~\ref{fig:contribution_HS3}. In the AstroR EGMF model, there is no significant contribution of additional sources for light and intermediate nuclei (p, He, and N). The contribution of Vir~A and For~A depend strongly on the EGMF model and ejected nuclei, being more important in the case of heavy nuclei (Si and Fe).

\section{Results II: Dipole generated by AGN\lowercase{s} and/or SBG\lowercase{s}}
\label{sec:dipoles}

\begin{figure}
  \centering
  \includegraphics[width=1.0\columnwidth]{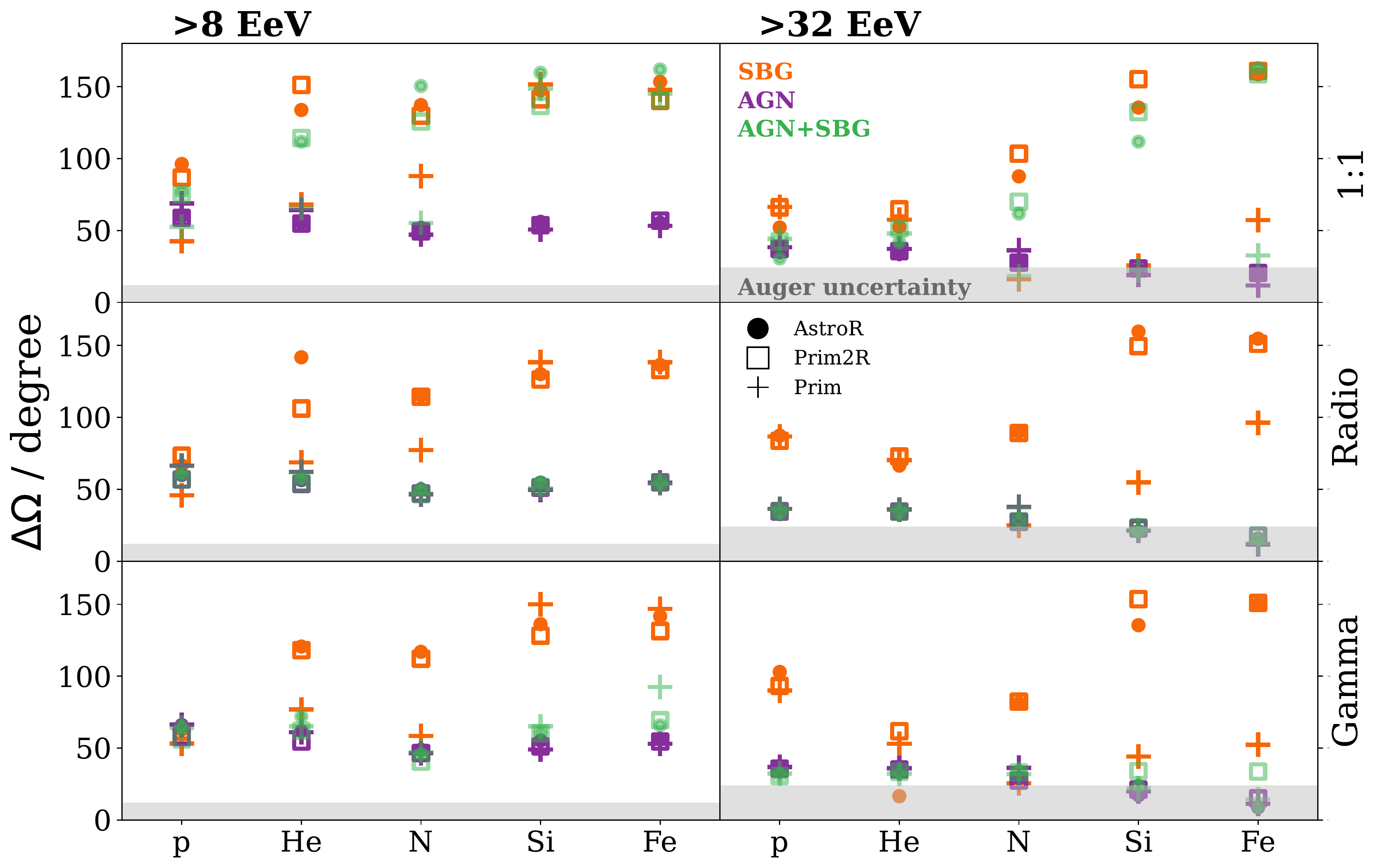}
  %\caption{Dipole direction generated by events with energy above 32~EeV if SBGs and AGNs are considered sources. The elements in the figure are the same of figure~\ref{fig:dip_8EeV_SBG-AGN}.}
  \caption{Angular aperture ($\Delta \Omega$) between the simulated dipole and the dipole direction measured by the Pierre Auger Observatory~\citep{deAlmeida:20212Z}. The data is presented as a function of the composition injected by the source (p, He, N, Si, and Fe). The left panel show the results for events with energy $>8$~EeV, while the right for $>32$~EeV. The top, center and bottom panels correspond to different luminosity proxies for the UHECR: identifical (1:1, top), radio (center), and gamma (bottom). The colors represent different sources class (SBG, AGN, or both), while the symbols represent different EGMF models (astroR, Prim2R, and Prim). The gray region corresponds to the uncertainty of the Pierre Auger data~\citep{deAlmeida:20212Z} to the different energy ranges considered.}
  \label{fig:dip:delta}
\end{figure}

\begin{figure}
  \centering
  \includegraphics[width=1.0\columnwidth]{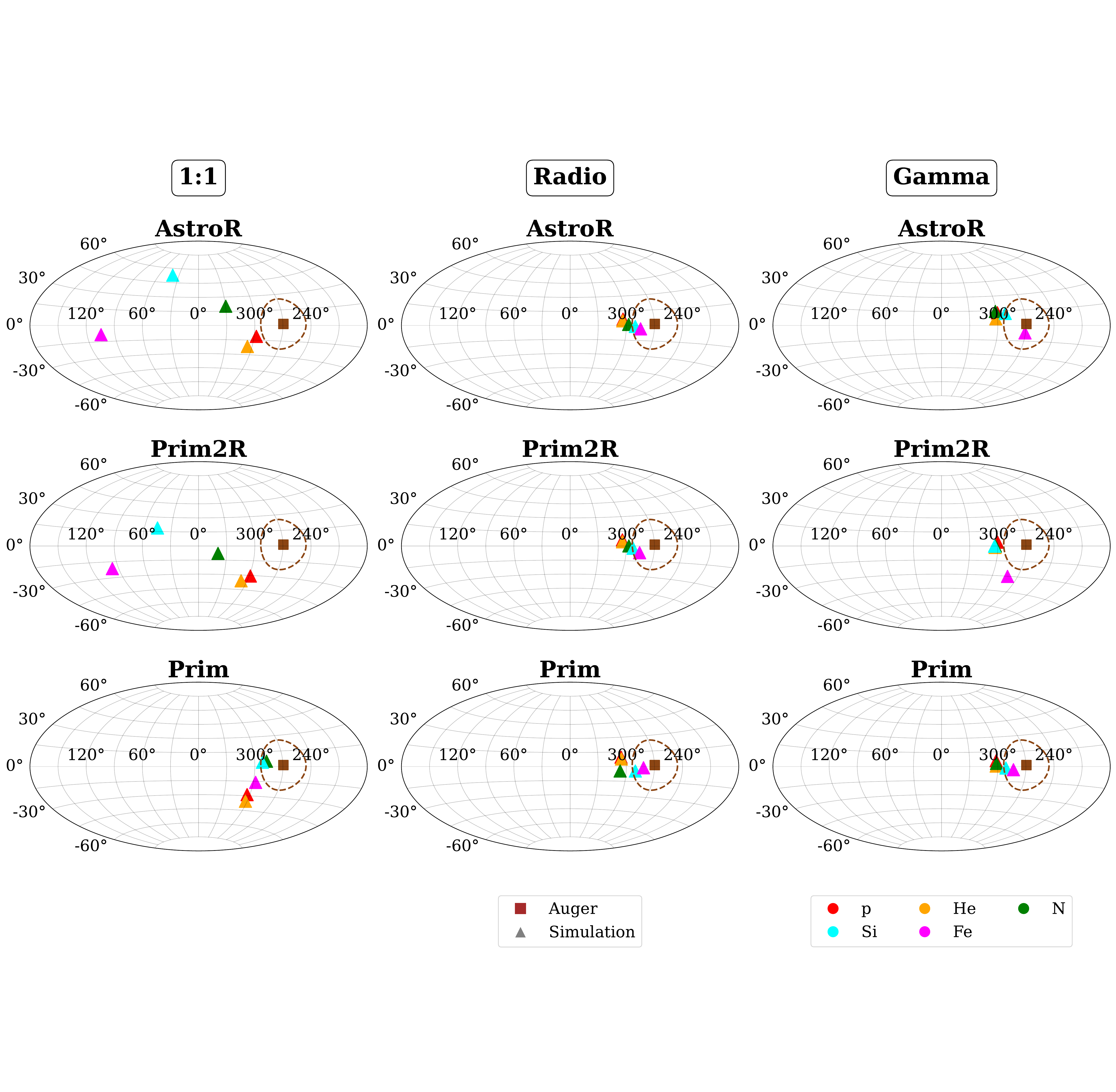}
  %\caption{Dipole direction generated by events with energy above 32~EeV if SBGs and AGNs are considered sources. The elements in the figure are the same of figure~\ref{fig:dip_8EeV_SBG-AGN}.}
  \caption{Dipole direction generated by events with energy above 32~EeV if SBGs and AGNs are considered sources. Each column presents different UHECR luminosities proxies: 1:1, Radio, and Gamma. Each line are showns different EGMF models: AstroR, Prim2R and Prim. The nuclei emitted by the source is mapped by colors: proton (p, red), helium (He, orange), nitrogen (N, green), silicon (Si, cyan), and iron (Fe, magenta). The brown square presents the dipole direction measured by the Pierre Auger Observatory~\citep{deAlmeida:20212Z}. The map is in galactic coordinates.}
  \label{fig:dip_32EeV_SBG-AGN}
\end{figure}

The dipole direction generated by the simulated events was evaluated using the methodology proposed by~\citet{aublin2005generalised}. The partial-sky coverage of the Pierre Auger Observatory was used. Different scenarios were considered:
\begin{itemize}
    \item Sources: only SBG, only AGN, or both SBG and AGN;
    \item UHECR luminosity: equal luminosity (1:1), scaling with $L_{radio}$, and scaling with $L_{\gamma}$;
    \item EGMF models: AstroR, Prim2R, and Prim;
    \item Nuclei injected by the source: proton, He, N, Si or Fe;
    \item Energy range of the events at Earth: above 8~EeV and above 32~EeV.
\end{itemize}
In each scenario, we compare the simulated results with the Pierre Auger Observatory data~\citep{deAlmeida:20212Z}. Figure \ref{fig:dip:delta} shows the angular aperture ($\Delta \Omega$) between the simulated dipole and the dipole direction measured by the Pierre Auger Observatory considering that only AGN, only SBGs, or both AGNs and SBGs are sources of UHECR. Figure \ref{fig:dip_32EeV_SBG-AGN} show the dipole direction in sky maps corresponding to the case in which AGNs and SBGs are sources, and events with energy above $32$~EeV are considered.

\subsection{Dipole generated by AGNs and SBGs}

%Figures \ref{fig:dip_8EeV_SBG-AGN} and \ref{fig:dip_32EeV_SBG-AGN} show the dipole evaluated for events with energies above 8~EeV and 32~EeV, respectively, when both SBGs and AGNs are sources of UHECR.
When both AGNs and SBGs are sources of UHECR, the direction of the simulated dipole is highly dependent on the UHECR luminosity proxies and the injected nuclei for both energy ranges. The dependence on the composition is more extreme in the case of equal UHECR luminosity of the sources. In the case of UHECR luminosity scaling with the radio luminosity, the dipole is dictated by Cen~A. When the UHECR luminosity scales with the gamma luminosity, the dipole is dictated by Cen~A, M82, and NGC4945.

For the energies above 8 EeV, no combination of EGMF model, UHECR luminosity proxies and the inject nuclei could reproduce the dipole direction measured by the Pierre Auger Observatory.

For the energies above 32 EeV, if the UHECR luminosity is considered proportional to the radio luminosity, the dipole direction measured by the Pierre Auger Observatory could be described for most of the injected nuclei and for all EGMF models. When the UHECR luminosity was considered proportional to the gamma luminosity the measured dipole direction could be described in some cases when heavy nuclei were injected.

\subsection{Dipole generated only by SBGs}

%Figures \ref{fig:dip_8EeV_SBG} and \ref{fig:dip_32EeV_SBG} show the dipole evaluated for events with energies above 8~EeV and 32~EeV, respectively, when the SBGs are UHECR sources.
When only SBGs are sources of UHECR, no combination of EGMF model, UHECR luminosity proxies and the inject nuclei could reproduce the dipole direction measured by the Pierre Auger Observatory, independently of the energy range considered. The dipole direction is highly dependent on the nuclei injected by the source, on the EGMF, and on the UHECR luminosity.

\subsection{Dipole generated only by AGNs}

%Figures \ref{fig:dip_8EeV_AGN} and \ref{fig:dip_32EeV_AGN} show the dipole evaluated for events with energies above 8~EeV and 32~EeV, respectively, when only Cen~A, Vir~A, and For~A are UHECR sources.
When only Cen~A, Vir~A, and For~A are UHECR sources, the dipole direction does not change significantly for the UHECR luminosity proxies considered here ($<17^\circ$ for energies above 8~EeV; and $<11^\circ$ for energies above 32~EeV). The dipole direction is dictated by Cen~A for both energy ranges ($> 8$ and $>32$ EeV) considered. As Cen~A is a very close source, the dipole direction is not greatly affected by the EGMF model ($<11^\circ$ for both energy ranges). In the case in which all the scenarios are compared the dipole direction changes less than $31^\circ$.

For the energies above 8~EeV, no combination of EGMF model, UHECR luminosity proxies and the inject nuclei could reproduce the dipole direction measured by the Pierre Auger Observatory.

For the energies above 32 EeV, most combinations of EGMF model, UHECR luminosity proxies and the inject nuclei could reproduce the dipole direction measured by the Pierre Auger Observatory within the uncertainties. The agreement between data and simulation does not depend on the EGMF model and UHECR luminosity proxies. The agreement between data and simulation improves with heavier nuclei.

\subsubsection{Searching the AGN luminosity which best describes the dipole direction}

The results presented in the previous section suggest that Cen~A, Vir A, and For~A can generate events to reproduce the measured direction of the dipole above 32~EeV. In this section, we allow the UHECR luminosity to vary and search for the best combination of UHECR flux coming Cen~A, Vir~A, and For~A which reproduces the measured direction of the dipole above 32~EeV. We varied the relative contribution of Vir~A and For~A in relation to Cen~A by a factor going from $10^{-2}$ to $10^2$ and calculated the dipole direction.

\begin{figure}
  \centering
  \includegraphics[width=1.0\columnwidth]{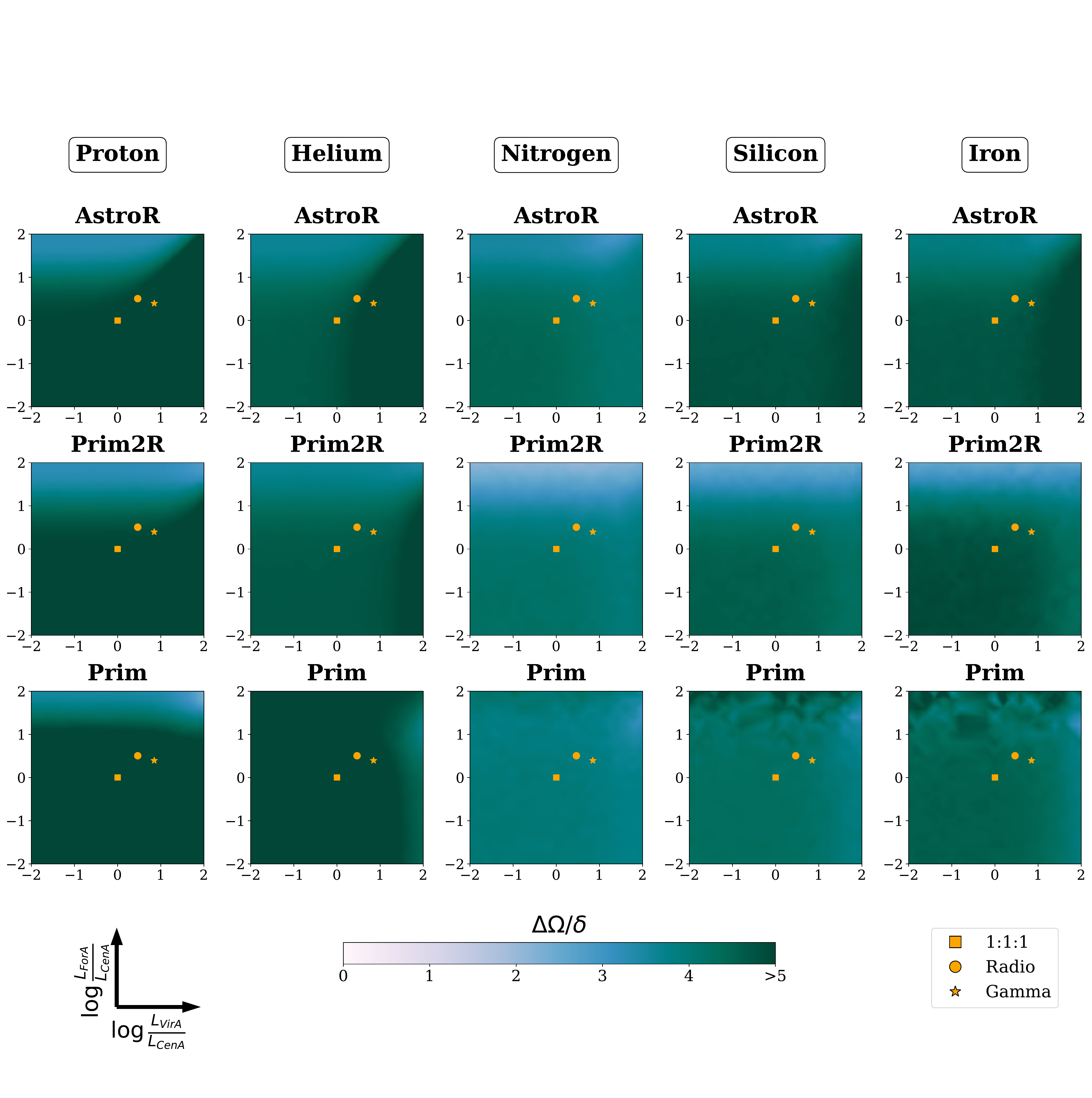}
  \caption{Normalized angular distance ($\frac{\Delta \Omega}{\delta}$) between simulated dipole direction and the direction of the dipole measured by the Pierre Auger Observatory~\citep{deAlmeida:20212Z} of events arriving at Earth with energy above 8~EeV. The three luminosity proxies considered in the previous sections are showed by the square (1:1), circle (Radio), and star (Gamma).}
  \label{fig:dip-AGN-proportions-8EeV}
\end{figure}

\begin{figure}
  \centering
  \includegraphics[width=1.0\columnwidth]{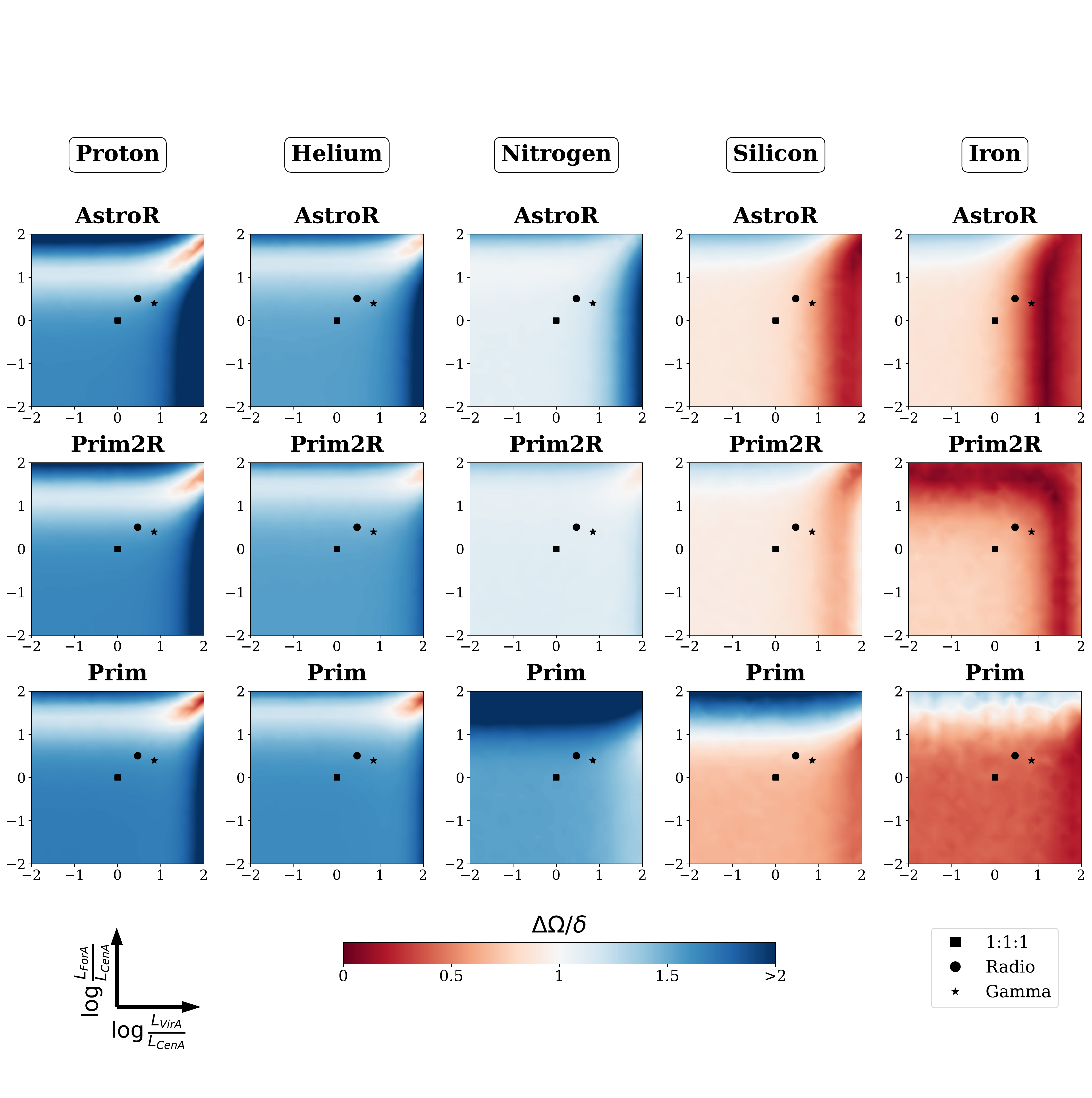}
  \caption{Normalized angular distance ($\frac{\Delta \Omega}{\delta}$) between simulated dipole direction and the direction of the dipole measured by the Pierre Auger Observatory~\citep{deAlmeida:20212Z} of events arriving at Earth with energy above 32~EeV. The angular distance is normalized by the uncertainty (${\delta}$) in the directon of the dipole reconstructed by The Pierre Auger Collaboration~\citep{deAlmeida:20212Z}. The three luminosity proxies considered in the previous sections are showed by the square (1:1), circle (Radio), and star (Gamma).}
  \label{fig:dip-AGN-proportions-32EeV}
\end{figure}

Figures~\ref{fig:dip-AGN-proportions-8EeV} and~\ref{fig:dip-AGN-proportions-32EeV} show the angular distance ($\Delta \Omega$) between the simulated dipole and the measure dipole as a function of the relative contribution of Vir~A ($L_{VirA}$) and For~A ($L_{ForA}$) in relation to Cen~A ($L_{CenA}$). The color code shows the angular distance in units of $\delta$, the uncertainty on the dipole direction determined by the Pierre Auger Collaboration. If the simulated dipole is inside (outside) the uncertainty of the measure dipole then $\frac{\Delta \Omega}{\delta} < 1$ ($>1$). The location of the three luminosity proxies considered in the previous sections is shown by the square (1:1:1), circle (Radio), and star (Gamma).

For the energy range above 8~EeV, the measured dipole could not be described by any combination of EGMF model, UHECR luminosity proxies, inject nuclei, and values of $L_{ForA}/L_{CenA}$ and $L_{VirA}/L_{CenA}$ tested here.

For the energy range above 32~EeV, the dipole can be reproduced for a large range of values of $L_{ForA}/L_{CenA}$ and $L_{VirA}/L_{CenA}$ if Silicon and Iron are injected for all EGMF models. If light nuclei are injected (p and He), there is a very narrow range of  $L_{ForA}/L_{CenA}$ and $L_{VirA}/L_{CenA}$ ($\sim 10-100$) for which the measured dipole could be described. If a intermediary nuclei (N) is considered, the simulated dipole can describe the data only with Prim2R EGMF model and with $L_{VirA} \sim L_{ForA} \geq 25 L_{CenA}$.

\section{Conclusion}
\label{sec:conclusion}

We investigated the measured anisotropic signal in UHECR arrival direction, namely the dipole and three hotspots, under the assumption that nearby AGNs and SBGs are the sources of these features. We studied different UHECR luminosity proxies (1:1, Radio, and Gamma), injection of several nuclei (p, He, N, Si, and Fe), and EGMF models (AstroR, Prim, and Prim2R). The importance of considering EGMF models in the study of UHECR anisotropy was once again confirmed~\citep{SIGL2004224,Tanco2001,Medina_Tanco_1998,1995ApJ...455L..21L,PhysRevD.68.043002,de_Oliveira_2022}. The dipole and the hotspots can not be interpreted without the consideration of structured EGMF models as the ones used here. Beyond the previous results, we showed here that the Prim2R EGMF model suppresses the flux arriving at Earth from several possible sources: NGC~5055, NGC~3628, NGC~3627, NGC~4631, M51, NGC~3556, NGC~3079, and Vir~A. The Prim EGMF model suppresses the flux of these sources in addition to M83, NGC~6946, and NGC~660. Prim EGMF model enhancements the flux of M82, NGC~4945, NGC~2903, NGC~2146, and For~A. This relative suppression/enhancement effect illustrates the importance of considering structured EGMF models in UHECR studies.

Our results suggest the hotspots measured by the Pierre Auger Observatory (HS1 and HS2) are dominated by nearby AGNs (Cen~A and For~A) emitting intermediate to heavy primaries~(figures \ref{fig:contribution_HS1} and \ref{fig:contribution_HS2}). Some SBGs (M83, NGC~253, and NGC~1068) also contribute with less importance. The HS3 is mostly populated by SBGs (NGC~2903 and NGC~3019) with a smaller contribution of Vir~A, however the results are more dependent on the EGMF model and injected nuclei.

The measured dipole for energies above 8 EeV cannot be described by nearby SBG only, neither by nearby AGN only, nor by nearby AGNs and SBGs combined. The measured dipole for energies above 32 EeV can be described by nearby AGN only and cannot be described by SBG only neither by AGNs and SBGs combined.

The astrophysical model emerging from these results requires UHECR production in AGNs and SBGs. The flux on Earth would be dominated by AGNs emitting intermediate to heavy primaries. This would explain the hotspots HS1 and HS2, is consistent with composition-sensitive measurements~\cite{PhysRevD.96.122003} and anisotropy beyond 32 EeV measured by the Pierre Auger Observatory. SBGs would contribute to the flux on Earth at a moderate percentage which is needed to explain HS3 measured by the Telescope Array Experiment.

The dipole measured for energies above 8 EeV can not be explained by any combination of AGNs and SBGs within the scenarios explored here, including three EGMF models, three UHECRs luminosity proxies, and inject nuclei from proton to iron. The dipole for energies above 8 EeV would need an extra source component, beyond local AGNs and SBGs, to be explained. The need for an extra component has been also advocated by studies of the energy spectrum~\cite{bib:hillas, Peixoto_2015}. Galactic sources have been considered, such as hyper-novae~\cite{bib:paczynski,bib:wang}, young neutron star winds~\cite{Blasi_2000} and magnetars~\cite{1475-7516-2014-10-020}. In these models, the injected composition would need to be heavier nuclei otherwise the maximum energy leaving the source would not exceed $10^{18}$ eV. The anisotropies generated in these galactic scenarios have not been studied. Another possibility would be a further ($>20$ Mpc) anomalous source for which the injected particles would arrive on Earth with energy around the photon-pion production threshold. Cygnus A can be one of these sources~\cite{eichmann2018ultra}, at a distance of about 250~Mpc (inside the magnetic horizon) and a radio luminosity of $1.6\times10^{44}$~erg/s~\cite{van2012radio}, it could contribute with intermediate mass of energy below $10^{19.5}$~eV. The relative large abundancy of SBG in respect to AGNs could generate more HS than the ones current measured. This was not investigated in this paper in which we focus on the explanation of the measured anisotropy signal.

\section*{Acknowledgments}
CO and VdS acknowledge FAPESP Project 2019/10151-2 and 2020/15453-4, 2021/01089-1. The authors acknowledge the National Laboratory for Scientific Computing (LNCC/MCTI,  Brazil) for providing HPC resources of the SDumont supercomputer (http://sdumont.lncc.br). VdS acknowledges CNPq. This study was financed in part by the Coordenação de Aperfeiçoamento de Pessoal de Nível Superior - Brasil (CAPES) - Finance Code 001.

\bibliographystyle{unsrtnat}
\bibliography{main.bib}

\begin{thebibliography}{85}
\providecommand{\natexlab}[1]{#1}
\providecommand{\url}[1]{\texttt{#1}}
\expandafter\ifx\csname urlstyle\endcsname\relax
  \providecommand{\doi}[1]{doi: #1}\else
  \providecommand{\doi}{doi: \begingroup \urlstyle{rm}\Url}\fi

\bibitem[Linsley(1963)]{PhysRevLett.10.146}
John Linsley.
\newblock Evidence for a primary cosmic-ray particle with energy ${10}^{20}$
  ev.
\newblock \emph{Phys. Rev. Lett.}, 10:\penalty0 146--148, Feb 1963.
\newblock \doi{10.1103/PhysRevLett.10.146}.
\newblock URL \url{https://link.aps.org/doi/10.1103/PhysRevLett.10.146}.

\bibitem[Nagano and Watson(2000)]{RevModPhys.72.689}
M.~Nagano and A.~A. Watson.
\newblock Observations and implications of the ultrahigh-energy cosmic rays.
\newblock \emph{Rev. Mod. Phys.}, 72:\penalty0 689--732, Jul 2000.
\newblock \doi{10.1103/RevModPhys.72.689}.
\newblock URL \url{https://link.aps.org/doi/10.1103/RevModPhys.72.689}.

\bibitem[Alves~Batista et~al.(2019)Alves~Batista, Biteau, Bustamante, Dolag,
  Engel, Fang, Kampert, Kostunin, Mostafa, Murase, et~al.]{alves2019open}
Rafael Alves~Batista, Jonathan Biteau, Mauricio Bustamante, Klaus Dolag, Ralph
  Engel, Ke~Fang, Karl-Heinz Kampert, Dmitriy Kostunin, Miguel Mostafa, Kohta
  Murase, et~al.
\newblock Open questions in cosmic-ray research at ultrahigh energies.
\newblock \emph{Frontiers in Astronomy and Space Sciences}, 6:\penalty0 23,
  2019.

\bibitem[Hillas(1984)]{hillas_1984}
A.~M. Hillas.
\newblock The origin of ultra-high-energy cosmic rays.
\newblock \emph{Annual Review of Astronomy and Astrophysics}, 22\penalty0
  (1):\penalty0 425--444, 1984.
\newblock \doi{10.1146/annurev.aa.22.090184.002233}.

\bibitem[Murase and Fukugita(2019)]{PhysRevD.99.063012}
Kohta Murase and Masataka Fukugita.
\newblock Energetics of high-energy cosmic radiations.
\newblock \emph{Phys. Rev. D}, 99:\penalty0 063012, Mar 2019.
\newblock \doi{10.1103/PhysRevD.99.063012}.
\newblock URL \url{https://link.aps.org/doi/10.1103/PhysRevD.99.063012}.

\bibitem[Ginzburg and Syrovatskii(1963)]{ginzburg1963cosmic}
VL~Ginzburg and SI~Syrovatskii.
\newblock Cosmic rays in metagalactic space.
\newblock \emph{Soviet Astronomy}, 7:\penalty0 357, 1963.

\bibitem[Caprioli(2015)]{Caprioli_2015}
Damiano Caprioli.
\newblock {\textquotedblleft}{ESPRESSO}{\textquotedblright} {ACCELERATION} {OF}
  {ULTRA}-{HIGH}-{ENERGY} {COSMIC} {RAYS}.
\newblock \emph{The Astrophysical Journal}, 811\penalty0 (2):\penalty0 L38, sep
  2015.
\newblock \doi{10.1088/2041-8205/811/2/l38}.
\newblock URL \url{https://doi.org/10.1088/2041-8205/811/2/l38}.

\bibitem[Liu et~al.(2017)Liu, Rieger, and Aharonian]{liu2017particle}
Ruo-Yu Liu, Frank~M Rieger, and Felix~A Aharonian.
\newblock Particle acceleration in mildly relativistic shearing flows: The
  interplay of systematic and stochastic effects, and the origin of the
  extended high-energy emission in {AGN} jets.
\newblock \emph{The Astrophysical Journal}, 842\penalty0 (1):\penalty0 39,
  2017.

\bibitem[Kimura et~al.(2018)Kimura, Murase, and Zhang]{PhysRevD.97.023026}
Shigeo~S. Kimura, Kohta Murase, and B.~Theodore Zhang.
\newblock Ultrahigh-energy cosmic-ray nuclei from black hole jets: Recycling
  galactic cosmic rays through shear acceleration.
\newblock \emph{Phys. Rev. D}, 97:\penalty0 023026, Jan 2018.
\newblock \doi{10.1103/PhysRevD.97.023026}.
\newblock URL \url{https://link.aps.org/doi/10.1103/PhysRevD.97.023026}.

\bibitem[Eichmann et~al.(2018)Eichmann, Rachen, Merten, van Vliet, and
  Tjus]{eichmann2018ultra}
Bj{\"o}rn Eichmann, J{\"o}rg~P Rachen, Lukas Merten, Arjen van Vliet, and
  J~Becker Tjus.
\newblock Ultra-high-energy cosmic rays from radio galaxies.
\newblock \emph{Journal of Cosmology and Astroparticle Physics}, 2018\penalty0
  (02):\penalty0 036, 2018.

\bibitem[Matthews et~al.(2019)Matthews, Bell, Blundell, and
  Araudo]{matthews2019ultrahigh}
James~H Matthews, Anthony~R Bell, Katherine~M Blundell, and Anabella~T Araudo.
\newblock Ultrahigh energy cosmic rays from shocks in the lobes of powerful
  radio galaxies.
\newblock \emph{Monthly Notices of the Royal Astronomical Society},
  482\penalty0 (4):\penalty0 4303--4321, 2019.

\bibitem[Eichmann(2019)]{eichmann2019high}
Bj{\"o}rn Eichmann.
\newblock High energy cosmic rays from {F}anaroff-{R}iley radio galaxies.
\newblock \emph{Journal of Cosmology and Astroparticle Physics}, 2019\penalty0
  (05):\penalty0 009, 2019.

\bibitem[Dermer et~al.(2009)Dermer, Razzaque, Finke, and
  Atoyan]{dermer2009ultra}
CD~Dermer, S~Razzaque, JD~Finke, and A~Atoyan.
\newblock Ultra-high-energy cosmic rays from black hole jets of radio galaxies.
\newblock \emph{New Journal of Physics}, 11\penalty0 (6):\penalty0 065016,
  2009.

\bibitem[van Velzen et~al.(2012)van Velzen, Falcke, Schellart,
  Nierstenh{\"o}fer, and Kampert]{van2012radio}
Sjoert van Velzen, Heino Falcke, Pim Schellart, Nils Nierstenh{\"o}fer, and
  Karl-Heinz Kampert.
\newblock Radio galaxies of the local universe: all-sky catalog, luminosity
  functions, and clustering.
\newblock \emph{Astronomy \& Astrophysics}, 544:\penalty0 A18, 2012.

\bibitem[{Cavagnolo} et~al.(2010){Cavagnolo}, {McNamara}, {Nulsen}, {Carilli},
  {Jones}, and {B{\^\i}rzan}]{cavagnolo2010}
K.~W. {Cavagnolo}, B.~R. {McNamara}, P.~E.~J. {Nulsen}, C.~L. {Carilli},
  C.~{Jones}, and L.~{B{\^\i}rzan}.
\newblock {A Relationship Between {AGN} Jet Power and Radio Power}.
\newblock \emph{The Astrophysical Journal}, 720\penalty0 (2):\penalty0
  1066--1072, Sep 2010.
\newblock \doi{10.1088/0004-637X/720/2/1066}.

\bibitem[Anchordoqui et~al.(1999)Anchordoqui, Romero, and
  Combi]{PhysRevD.60.103001}
L.~A. Anchordoqui, G.~E. Romero, and J.~A. Combi.
\newblock Heavy nuclei at the end of the cosmic-ray spectrum?
\newblock \emph{Phys. Rev. D}, 60:\penalty0 103001, Oct 1999.
\newblock \doi{10.1103/PhysRevD.60.103001}.
\newblock URL \url{https://link.aps.org/doi/10.1103/PhysRevD.60.103001}.

\bibitem[Anchordoqui(2018)]{PhysRevD.97.063010}
Luis~Alfredo Anchordoqui.
\newblock Acceleration of ultrahigh-energy cosmic rays in starburst superwinds.
\newblock \emph{Phys. Rev. D}, 97:\penalty0 063010, Mar 2018.
\newblock \doi{10.1103/PhysRevD.97.063010}.
\newblock URL \url{https://link.aps.org/doi/10.1103/PhysRevD.97.063010}.

\bibitem[{Romero, G. E.} et~al.(2018){Romero, G. E.}, {M\"uller, A. L.}, and
  {Roth, M.}]{romero2018_sbgwinds}
{Romero, G. E.}, {M\"uller, A. L.}, and {Roth, M.}
\newblock Particle acceleration in the superwinds of starburst galaxies.
\newblock \emph{A\&A}, 616:\penalty0 A57, 2018.
\newblock \doi{10.1051/0004-6361/201832666}.
\newblock URL \url{https://doi.org/10.1051/0004-6361/201832666}.

\bibitem[Matthews et~al.(2018)Matthews, Bell, Blundell, and
  Araudo]{matthews2018fornax}
James~H Matthews, Anthony~R Bell, Katherine~M Blundell, and Anabella~T Araudo.
\newblock Fornax {A}, {C}entaurus {A}, and other radio galaxies as sources of
  ultrahigh energy cosmic rays.
\newblock \emph{Monthly Notices of the Royal Astronomical Society: Letters},
  479\penalty0 (1):\penalty0 L76--L80, 2018.

\bibitem[Kotera and Lemoine(2008)]{PhysRevD.77.123003}
Kumiko Kotera and Martin Lemoine.
\newblock Optical depth of the universe to ultrahigh energy cosmic ray
  scattering in the magnetized large scale structure.
\newblock \emph{Phys. Rev. D}, 77:\penalty0 123003, Jun 2008.
\newblock \doi{10.1103/PhysRevD.77.123003}.
\newblock URL \url{https://link.aps.org/doi/10.1103/PhysRevD.77.123003}.

\bibitem[Bell and Matthews(2022)]{10.1093/mnras/stac031}
A~R Bell and J~H Matthews.
\newblock {Echoes of the past: ultra-high-energy cosmic rays accelerated by
  radio galaxies, scattered by starburst galaxies}.
\newblock \emph{Monthly Notices of the Royal Astronomical Society},
  511\penalty0 (1):\penalty0 448--456, 01 2022.
\newblock ISSN 0035-8711.
\newblock \doi{10.1093/mnras/stac031}.
\newblock URL \url{https://doi.org/10.1093/mnras/stac031}.

\bibitem[Zhang and Murase(2019)]{PhysRevD.100.103004}
B.~Theodore Zhang and Kohta Murase.
\newblock Ultrahigh-energy cosmic-ray nuclei and neutrinos from engine-driven
  supernovae.
\newblock \emph{Phys. Rev. D}, 100:\penalty0 103004, Nov 2019.
\newblock \doi{10.1103/PhysRevD.100.103004}.
\newblock URL \url{https://link.aps.org/doi/10.1103/PhysRevD.100.103004}.

\bibitem[Zhang et~al.(2018)Zhang, Murase, Kimura, Horiuchi, and
  M\'esz\'aros]{PhysRevD.97.083010}
B.~Theodore Zhang, Kohta Murase, Shigeo~S. Kimura, Shunsaku Horiuchi, and Peter
  M\'esz\'aros.
\newblock Low-luminosity gamma-ray bursts as the sources of ultrahigh-energy
  cosmic ray nuclei.
\newblock \emph{Phys. Rev. D}, 97:\penalty0 083010, Apr 2018.
\newblock \doi{10.1103/PhysRevD.97.083010}.
\newblock URL \url{https://link.aps.org/doi/10.1103/PhysRevD.97.083010}.

\bibitem[Biermann et~al.(2016)Biermann, Caramete, Fraschetti, Gergely, Harms,
  Kun, Lundquist, Meli, Nath, Seo, Stanev, and Tjus]{arxiv.1610.00944}
Peter~L. Biermann, Laurentiu~I. Caramete, Federico Fraschetti, Laszlo~A.
  Gergely, Benjamin~C. Harms, Emma Kun, Jon~Paul Lundquist, Athina Meli,
  Biman~B. Nath, Eun-Suk Seo, Todor Stanev, and Julia~Becker Tjus.
\newblock The nature and origin of ultra-high energy cosmic ray particles,
  2016.
\newblock URL \url{https://arxiv.org/abs/1610.00944}.

\bibitem[{The Pierre Auger Collaboration}(2015)]{bib:auger}
{The Pierre Auger Collaboration}.
\newblock {The Pierre Auger Cosmic Ray Observatory}.
\newblock \emph{Nuclear Instruments and Methods in Physics Research Section A:
  Accelerators, Spectrometers, Detectors and Associated Equipment},
  798:\penalty0 172--213, 2015.
\newblock ISSN 0168-9002.
\newblock \doi{https://doi.org/10.1016/j.nima.2015.06.058}.
\newblock URL
  \url{https://www.sciencedirect.com/science/article/pii/S0168900215008086}.

\bibitem[{T. Abu-Zayyad et al. (The Telescope Array
  Collaboration)}(2012)]{bib:ta}
{T. Abu-Zayyad et al. (The Telescope Array Collaboration)}.
\newblock The surface detector array of the {Telescope Array} experiment.
\newblock \emph{Nuclear Instruments and Methods in Physics Research Section A:
  Accelerators, Spectrometers, Detectors and Associated Equipment},
  689:\penalty0 87–97, Oct 2012.
\newblock ISSN 0168-9002.
\newblock \doi{10.1016/j.nima.2012.05.079}.
\newblock URL \url{http://dx.doi.org/10.1016/j.nima.2012.05.079}.

\bibitem[{The Pierre Auger Collaboration}(2017)]{1266}
{The Pierre Auger Collaboration}.
\newblock Observation of a large-scale anisotropy in the arrival directions of
  cosmic rays above 8 {\texttimes} 10$^{18}$ {eV}.
\newblock \emph{Science}, 357\penalty0 (6357):\penalty0 1266--1270, 2017.
\newblock ISSN 0036-8075.
\newblock \doi{10.1126/science.aan4338}.
\newblock URL \url{https://science.sciencemag.org/content/357/6357/1266}.

\bibitem[{The Pierre Auger Collaboration}(2018{\natexlab{a}})]{Aab_2018}
{The Pierre Auger Collaboration}.
\newblock Large-scale cosmic-ray anisotropies above 4 {EeV} measured by the
  {Pierre Auger Observatory}.
\newblock \emph{The Astrophysical Journal}, 868\penalty0 (1):\penalty0 4, nov
  2018{\natexlab{a}}.
\newblock \doi{10.3847/1538-4357/aae689}.
\newblock URL \url{https://doi.org/10.3847/1538-4357/aae689}.

\bibitem[Aab et~al.(2020{\natexlab{a}})Aab, Abreu, Aglietta, Albuquerque,
  Albury, Allekotte, Almela, Castillo, Alvarez-Mu{\~{n}}iz, Anastasi,
  Anchordoqui, Andrada, Andringa, Aramo, Ferreira, Asorey, Assis, Avila,
  Badescu, Bakalova, Balaceanu, Barbato, Luz, Becker, Bellido, Berat, Bertaina,
  Bertou, Biermann, Bister, Biteau, Blanco, Blazek, Bleve,
  Boh{\'{a}}{\v{c}}ov{\'{a}}, Boncioli, Bonifazi, Arbeletche, Borodai, Botti,
  Brack, Bretz, Briechle, Buchholz, Bueno, Buitink, Buscemi, Caballero-Mora,
  Caccianiga, Calcagni, Cancio, Canfora, Caracas, Carceller, Caruso,
  Castellina, Catalani, Cataldi, Cazon, Cerda, Chinellato, Choi, Chudoba,
  Chytka, Clay, Cerutti, Colalillo, Coleman, Coluccia, Concei{\c{c}}{\~{a}}o,
  Condorelli, Consolati, Contreras, Convenga, Covault, Dasso, Daumiller,
  Dawson, Day, de~Almeida, de~Jes{\'{u}}s, de~Jong, Mauro, de~Mello~Neto,
  Mitri, de~Oliveira, de~Oliveira~Franco, de~Souza, Debatin, del R{\'{\i}}o,
  Deligny, Dhital, Matteo, Castro, Dobrigkeit, D'Olivo, Dorosti, dos Anjos,
  Dova, Ebr, Engel, Epicoco, Erdmann, Escobar, Etchegoyen, Falcke, Farmer,
  Farrar, Fauth, Fazzini, Feldbusch, Fenu, Fick, Figueira,
  Filip{\v{c}}i{\v{c}}, Freire, Fujii, Fuster, Galea, Galelli, Garc{\'{\i}}a,
  Vegas, Gemmeke, Gesualdi, Gherghel-Lascu, Ghia, Giaccari, Giammarchi, Giller,
  Glombitza, Gobbi, Golup, Berisso, Vitale, Gongora, Gonz{\'{a}}lez, Goos,
  G{\'{o}}ra, Gorgi, Gottowik, Grubb, Guarino, Guedes, Guido, Hahn, Halliday,
  Hampel, Hansen, Harari, Harvey, Haungs, Hebbeker, Heck, Hill, Hojvat,
  Hörandel, Horvath, Hrabovsk{\'{y}}, Huege, Hulsman, Insolia, Isar, Johnsen,
  Jurysek, Kääpä, Kampert, Keilhauer, Kemp, Klages, Kleifges, Kleinfeller,
  Köpke, Mezek, Awad, Lago, LaHurd, Lang, de~Oliveira, Lenok,
  Letessier-Selvon, Lhenry-Yvon, Presti, Lopes, L{\'{o}}pez, Casado, Lorek,
  Luce, Lucero, Payeras, Malacari, Mancarella, Mandat, Manning, Manshanden,
  Mantsch, Mariazzi, Mari{\c{s}}, Marsella, Martello, Martinez, Bravo,
  Mastrodicasa, Mathes, Matthews, Matthiae, Mayotte, Mazur, Medina-Tanco, Melo,
  Menshikov, Merenda, Michal, Micheletti, Miramonti, Mockler, Mollerach,
  Montanet, Morello, Morlino, Mostaf{\'{a}}, Müller, Muller, Müller, Mussa,
  Muzio, Namasaka, Nellen, Niculescu-Oglinzanu, Niechciol, Nitz, Nosek,
  Novotny, No{\v{z}}ka, Nucita, N{\'{u}}{\~{n}}ez, Palatka, Pallotta, Panetta,
  Papenbreer, Parente, Parra, Pech, Pedreira, Pekala, Pelayo,
  Pe{\~{n}}a-Rodriguez, Pereira, Armand, Perlin, Perrone, Peters, Petrera,
  Pierog, Pimenta, Pirronello, Platino, Pont, Pothast, Privitera, Prouza,
  Puyleart, Querchfeld, Rautenberg, Ravignani, Reininghaus, Ridky, Riehn,
  Risse, Ristori, Rizi, de~Carvalho, Rojo, Roncoroni, Roth, Roulet, Rovero,
  Ruehl, Saffi, Saftoiu, Salamida, Salazar, Salina, Gomez, S{\'{a}}nchez,
  Santos, Santos, Sarazin, Sarmento, Sarmiento-Cano, Sato, Savina, Schäfer,
  Scherini, Schieler, Schimassek, Schimp, Schlüter, Schmidt, Scholten,
  Schov{\'{a}}nek, Schröder, Schröder, Sciutto, Scornavacche, Shellard, Sigl,
  Silli, Sima, {\v{S}}m{\'{\i}}da, Sommers, Soriano, Souchard, Squartini,
  Stadelmaier, Stanca, Stani{\v{c}}, Stasielak, Stassi, Streich,
  Su{\'{a}}rez-Dur{\'{a}}n, Sudholz, Suomijärvi, Supanitsky,
  {\v{S}}up{\'{\i}}k, Szadkowski, Taboada, Taborda, Tapia, Timmermans, Tobiska,
  Peixoto, Tom{\'{e}}, Elipe, Travaini, Travnicek, Trimarelli, Trini, Tueros,
  Ulrich, Unger, Urban, Vaclavek, Galicia, Vali{\~{n}}o, Valore, van Vliet,
  Varela, C{\'{a}}rdenas, V{\'{a}}squez-Ram{\'{\i}}rez, Veberi{\v{c}}, Ventura,
  Quispe, Verzi, Vicha, Villase{\~{n}}or, Vink, Vorobiov, Wahlberg, Watson,
  Weber, Weindl, Wiencke, Wilczy{\'{n}}ski, Winchen, Wirtz, Wittkowski,
  Wundheiler, Yushkov, Zas, Zavrtanik, Zavrtanik, Zehrer, Zepeda, Ziolkowski,
  and and]{Aab_2020_raAnisotropies}
A.~Aab, P.~Abreu, M.~Aglietta, I.~F.~M. Albuquerque, J.~M. Albury,
  I.~Allekotte, A.~Almela, J.~Alvarez Castillo, J.~Alvarez-Mu{\~{n}}iz, G.~A.
  Anastasi, L.~Anchordoqui, B.~Andrada, S.~Andringa, C.~Aramo,
  P.~R.~Ara{\'{u}}jo Ferreira, H.~Asorey, P.~Assis, G.~Avila, A.~M. Badescu,
  A.~Bakalova, A.~Balaceanu, F.~Barbato, R.~J.~Barreira Luz, K.~H. Becker,
  J.~A. Bellido, C.~Berat, M.~E. Bertaina, X.~Bertou, P.~L. Biermann,
  T.~Bister, J.~Biteau, A.~Blanco, J.~Blazek, C.~Bleve,
  M.~Boh{\'{a}}{\v{c}}ov{\'{a}}, D.~Boncioli, C.~Bonifazi, L.~Bonneau
  Arbeletche, N.~Borodai, A.~M. Botti, J.~Brack, T.~Bretz, F.~L. Briechle,
  P.~Buchholz, A.~Bueno, S.~Buitink, M.~Buscemi, K.~S. Caballero-Mora,
  L.~Caccianiga, L.~Calcagni, A.~Cancio, F.~Canfora, I.~Caracas, J.~M.
  Carceller, R.~Caruso, A.~Castellina, F.~Catalani, G.~Cataldi, L.~Cazon,
  M.~Cerda, J.~A. Chinellato, K.~Choi, J.~Chudoba, L.~Chytka, R.~W. Clay,
  A.~C.~Cobos Cerutti, R.~Colalillo, A.~Coleman, M.~R. Coluccia,
  R.~Concei{\c{c}}{\~{a}}o, A.~Condorelli, G.~Consolati, F.~Contreras,
  F.~Convenga, C.~E. Covault, S.~Dasso, K.~Daumiller, B.~R. Dawson, J.~A. Day,
  R.~M. de~Almeida, J.~de~Jes{\'{u}}s, S.~J. de~Jong, G.~De Mauro, J.~R.~T.
  de~Mello~Neto, I.~De Mitri, J.~de~Oliveira, D.~de~Oliveira~Franco,
  V.~de~Souza, J.~Debatin, M.~del R{\'{\i}}o, O.~Deligny, N.~Dhital, A.~Di
  Matteo, M.~L.~D{\'{\i}}az Castro, C.~Dobrigkeit, J.~C. D'Olivo, Q.~Dorosti,
  R.~C. dos Anjos, M.~T. Dova, J.~Ebr, R.~Engel, I.~Epicoco, M.~Erdmann, C.~O.
  Escobar, A.~Etchegoyen, H.~Falcke, J.~Farmer, G.~Farrar, A.~C. Fauth,
  N.~Fazzini, F.~Feldbusch, F.~Fenu, B.~Fick, J.~M. Figueira,
  A.~Filip{\v{c}}i{\v{c}}, M.~M. Freire, T.~Fujii, A.~Fuster, C.~Galea,
  C.~Galelli, B.~Garc{\'{\i}}a, A.~L.~Garcia Vegas, H.~Gemmeke, F.~Gesualdi,
  A.~Gherghel-Lascu, P.~L. Ghia, U.~Giaccari, M.~Giammarchi, M.~Giller,
  J.~Glombitza, F.~Gobbi, G.~Golup, M.~G{\'{o}}mez Berisso, P.~F.~G{\'{o}}mez
  Vitale, J.~P. Gongora, N.~Gonz{\'{a}}lez, I.~Goos, D.~G{\'{o}}ra, A.~Gorgi,
  M.~Gottowik, T.~D. Grubb, F.~Guarino, G.~P. Guedes, E.~Guido, S.~Hahn,
  R.~Halliday, M.~R. Hampel, P.~Hansen, D.~Harari, V.~M. Harvey, A.~Haungs,
  T.~Hebbeker, D.~Heck, G.~C. Hill, C.~Hojvat, J.~R. Hörandel, P.~Horvath,
  M.~Hrabovsk{\'{y}}, T.~Huege, J.~Hulsman, A.~Insolia, P.~G. Isar, J.~A.
  Johnsen, J.~Jurysek, A.~Kääpä, K.~H. Kampert, B.~Keilhauer, J.~Kemp, H.~O.
  Klages, M.~Kleifges, J.~Kleinfeller, M.~Köpke, G.~Kukec Mezek, A.~Kuotb
  Awad, B.~L. Lago, D.~LaHurd, R.~G. Lang, M.~A.~Leigui de~Oliveira, V.~Lenok,
  A.~Letessier-Selvon, I.~Lhenry-Yvon, D.~Lo Presti, L.~Lopes, R.~L{\'{o}}pez,
  A.~L{\'{o}}pez Casado, R.~Lorek, Q.~Luce, A.~Lucero, A.~Machado Payeras,
  M.~Malacari, G.~Mancarella, D.~Mandat, B.~C. Manning, J.~Manshanden,
  P.~Mantsch, A.~G. Mariazzi, I.~C. Mari{\c{s}}, G.~Marsella, D.~Martello,
  H.~Martinez, O.~Mart{\'{\i}}nez Bravo, M.~Mastrodicasa, H.~J. Mathes,
  J.~Matthews, G.~Matthiae, E.~Mayotte, P.~O. Mazur, G.~Medina-Tanco, D.~Melo,
  A.~Menshikov, K.-D. Merenda, S.~Michal, M.~I. Micheletti, L.~Miramonti,
  D.~Mockler, S.~Mollerach, F.~Montanet, C.~Morello, G.~Morlino,
  M.~Mostaf{\'{a}}, A.~L. Müller, M.~A. Muller, S.~Müller, R.~Mussa,
  M.~Muzio, W.~M. Namasaka, L.~Nellen, M.~Niculescu-Oglinzanu, M.~Niechciol,
  D.~Nitz, D.~Nosek, V.~Novotny, L.~No{\v{z}}ka, A~Nucita, L.~A.
  N{\'{u}}{\~{n}}ez, M.~Palatka, J.~Pallotta, M.~P. Panetta, P.~Papenbreer,
  G.~Parente, A.~Parra, M.~Pech, F.~Pedreira, J.~Pekala, R.~Pelayo,
  J.~Pe{\~{n}}a-Rodriguez, L.~A.~S. Pereira, J.~Perez Armand, M.~Perlin,
  L.~Perrone, C.~Peters, S.~Petrera, T.~Pierog, M.~Pimenta, V.~Pirronello,
  M.~Platino, B.~Pont, M.~Pothast, P.~Privitera, M.~Prouza, A.~Puyleart,
  S.~Querchfeld, J.~Rautenberg, D.~Ravignani, M.~Reininghaus, J.~Ridky,
  F.~Riehn, M.~Risse, P.~Ristori, V.~Rizi, W.~Rodrigues de~Carvalho,
  J.~Rodriguez Rojo, M.~J. Roncoroni, M.~Roth, E.~Roulet, A.~C. Rovero,
  P.~Ruehl, S.~J. Saffi, A.~Saftoiu, F.~Salamida, H.~Salazar, G.~Salina,
  J.~D.~Sanabria Gomez, F.~S{\'{a}}nchez, E.~M. Santos, E.~Santos, F.~Sarazin,
  R.~Sarmento, C.~Sarmiento-Cano, R.~Sato, P.~Savina, C.~Schäfer, V.~Scherini,
  H.~Schieler, M.~Schimassek, M.~Schimp, F.~Schlüter, D.~Schmidt, O.~Scholten,
  P.~Schov{\'{a}}nek, F.~G. Schröder, S.~Schröder, S.~J. Sciutto,
  M.~Scornavacche, R.~C. Shellard, G.~Sigl, G.~Silli, O.~Sima,
  R.~{\v{S}}m{\'{\i}}da, P.~Sommers, J.~F. Soriano, J.~Souchard, R.~Squartini,
  M.~Stadelmaier, D.~Stanca, S.~Stani{\v{c}}, J.~Stasielak, P.~Stassi,
  A.~Streich, M.~Su{\'{a}}rez-Dur{\'{a}}n, T.~Sudholz, T.~Suomijärvi, A.~D.
  Supanitsky, J.~{\v{S}}up{\'{\i}}k, Z.~Szadkowski, A.~Taboada, O.~A. Taborda,
  A.~Tapia, C.~Timmermans, P.~Tobiska, C.~J.~Todero Peixoto, B.~Tom{\'{e}},
  G.~Torralba Elipe, A.~Travaini, P.~Travnicek, C.~Trimarelli, M.~Trini,
  M.~Tueros, R.~Ulrich, M.~Unger, M.~Urban, L.~Vaclavek, J.~F.~Vald{\'{e}}s
  Galicia, I.~Vali{\~{n}}o, L.~Valore, A.~van Vliet, E.~Varela, B.~Vargas
  C{\'{a}}rdenas, A.~V{\'{a}}squez-Ram{\'{\i}}rez, D.~Veberi{\v{c}},
  C.~Ventura, I.~D.~Vergara Quispe, V.~Verzi, J.~Vicha, L.~Villase{\~{n}}or,
  J.~Vink, S.~Vorobiov, H.~Wahlberg, A.~A. Watson, M.~Weber, A.~Weindl,
  L.~Wiencke, H.~Wilczy{\'{n}}ski, T.~Winchen, M.~Wirtz, D.~Wittkowski,
  B.~Wundheiler, A.~Yushkov, E.~Zas, D.~Zavrtanik, M.~Zavrtanik, L.~Zehrer,
  A.~Zepeda, M.~Ziolkowski, and F.~Zuccarello and.
\newblock Cosmic-ray anisotropies in right ascension measured by the pierre
  auger observatory.
\newblock \emph{The Astrophysical Journal}, 891\penalty0 (2):\penalty0 142, mar
  2020{\natexlab{a}}.
\newblock \doi{10.3847/1538-4357/ab7236}.
\newblock URL \url{https://doi.org/10.3847/1538-4357/ab7236}.

\bibitem[{The Pierre Auger Collaboration}(2021)]{deAlmeida:20212Z}
{The Pierre Auger Collaboration}.
\newblock {Large-scale and multipolar anisotropies of cosmic rays detected at
  the Pierre Auger Observatory with energies above 4 EeV}.
\newblock In \emph{Proceedings of 37th International Cosmic Ray Conference
  {\textemdash} PoS(ICRC2021)}, volume 395, page 335, 2021.
\newblock \doi{10.22323/1.395.0335}.

\bibitem[Abbasi et~al.(2020)Abbasi, Abe, Abu-Zayyad, Allen, Azuma, Barcikowski,
  Belz, Bergman, Blake, Cady, Cheon, Chiba, Chikawa, di~Matteo, Fujii, Fujisue,
  Fujita, Fujiwara, Fukushima, Furlich, Hanlon, Hayashi, Hayashida, Hibino,
  Higuchi, Honda, Ikeda, Inadomi, Inoue, Ishii, Ishimori, Ito, Ivanov, Iwakura,
  Jeong, Jeong, Jui, Kadota, Kakimoto, Kalashev, Kasahara, Kasami, Kawai,
  Kawakami, Kawana, Kawata, Kido, Kim, Kim, Kim, Kim, Kim, Kishigami, Kuzmin,
  Kuznetsov, Kwon, Lee, Lubsandorzhiev, Lundquist, Machida, Matsumiya,
  Matsuyama, Matthews, Mayta, Minamino, Mukai, Myers, Nagataki, Nakai,
  Nakamura, Nakamura, Nakamura, Nakamura, Nonaka, Oda, Ogio, Ohnishi, Ohoka,
  Oku, Okuda, Omura, Ono, Onogi, Oshima, Ozawa, Park, Pshirkov, Remington,
  Rodriguez, Rubtsov, Ryu, Sagawa, Sahara, Saito, Sakaki, Sako, Sakurai, Sano,
  Seki, Sekino, Shah, Shibata, Shibata, Shimodaira, Shin, Shin, Smith,
  Sokolsky, Sone, Stokes, Stroman, Suzawa, Takagi, Takahashi, Takamura, Takeda,
  Takeishi, Taketa, Takita, Tameda, Tanaka, Tanaka, Tanaka, Tanoue, Thomas,
  Thomson, Tinyakov, Tkachev, Tokuno, Tomida, Troitsky, Tsunesada, Uchihori,
  Udo, Uehama, Urban, Wong, Yada, Yamamoto, Yamazaki, Yang, Yashiro, Yosei,
  Zhezher, and and]{Abbasi_2020_largescale_anisotropy}
R.~U. Abbasi, M.~Abe, T.~Abu-Zayyad, M.~Allen, R.~Azuma, E.~Barcikowski, J.~W.
  Belz, D.~R. Bergman, S.~A. Blake, R.~Cady, B.~G. Cheon, J.~Chiba, M.~Chikawa,
  A.~di~Matteo, T.~Fujii, K.~Fujisue, K.~Fujita, R.~Fujiwara, M.~Fukushima,
  G.~Furlich, W.~Hanlon, M.~Hayashi, N.~Hayashida, K.~Hibino, R.~Higuchi,
  K.~Honda, D.~Ikeda, T.~Inadomi, N.~Inoue, T.~Ishii, R.~Ishimori, H.~Ito,
  D.~Ivanov, H.~Iwakura, H.~M. Jeong, S.~Jeong, C.~C.~H. Jui, K.~Kadota,
  F.~Kakimoto, O.~Kalashev, K.~Kasahara, S.~Kasami, H.~Kawai, S.~Kawakami,
  S.~Kawana, K.~Kawata, E.~Kido, H.~B. Kim, J.~H. Kim, J.~H. Kim, M.~H. Kim,
  S.~W. Kim, S.~Kishigami, V.~Kuzmin, M.~Kuznetsov, Y.~J. Kwon, K.~H. Lee,
  B.~Lubsandorzhiev, J.~P. Lundquist, K.~Machida, H.~Matsumiya, T.~Matsuyama,
  J.~N. Matthews, R.~Mayta, M.~Minamino, K.~Mukai, I.~Myers, S.~Nagataki,
  K.~Nakai, R.~Nakamura, T.~Nakamura, Y.~Nakamura, Y.~Nakamura, T.~Nonaka,
  H.~Oda, S.~Ogio, M.~Ohnishi, H.~Ohoka, Y.~Oku, T.~Okuda, Y.~Omura, M.~Ono,
  R.~Onogi, A.~Oshima, S.~Ozawa, I.~H. Park, M.~S. Pshirkov, J.~Remington,
  D.~C. Rodriguez, G.~Rubtsov, D.~Ryu, H.~Sagawa, R.~Sahara, Y.~Saito,
  N.~Sakaki, T.~Sako, N.~Sakurai, K.~Sano, T.~Seki, K.~Sekino, P.~D. Shah,
  F.~Shibata, T.~Shibata, H.~Shimodaira, B.~K. Shin, H.~S. Shin, J.~D. Smith,
  P.~Sokolsky, N.~Sone, B.~T. Stokes, T.~A. Stroman, T.~Suzawa, Y.~Takagi,
  Y.~Takahashi, M.~Takamura, M.~Takeda, R.~Takeishi, A.~Taketa, M.~Takita,
  Y.~Tameda, H.~Tanaka, K.~Tanaka, M.~Tanaka, Y.~Tanoue, S.~B. Thomas, G.~B.
  Thomson, P.~Tinyakov, I.~Tkachev, H.~Tokuno, T.~Tomida, S.~Troitsky,
  Y.~Tsunesada, Y.~Uchihori, S.~Udo, T.~Uehama, F.~Urban, T.~Wong, K.~Yada,
  M.~Yamamoto, K.~Yamazaki, J.~Yang, K.~Yashiro, M.~Yosei, Y.~Zhezher, and
  Z.~Zundel and.
\newblock Search for large-scale anisotropy on arrival directions of
  ultra-high-energy cosmic rays observed with the telescope array experiment.
\newblock \emph{The Astrophysical Journal}, 898\penalty0 (2):\penalty0 L28, jul
  2020.
\newblock \doi{10.3847/2041-8213/aba0bc}.
\newblock URL \url{https://doi.org/10.3847/2041-8213/aba0bc}.

\bibitem[{The Pierre Auger
  Collaboration}(2018{\natexlab{b}})]{aab2018indication}
{The Pierre Auger Collaboration}.
\newblock An indication of anisotropy in arrival directions of
  ultra-high-energy cosmic rays through comparison to the flux pattern of
  extragalactic gamma-ray sources.
\newblock \emph{The Astrophysical Journal Letters}, 853\penalty0 (2):\penalty0
  L29, 2018{\natexlab{b}}.

\bibitem[de~Oliveira and de~Souza(2022)]{de_Oliveira_2022}
Cain{\~{a}} de~Oliveira and Vitor de~Souza.
\newblock Magnetically induced anisotropies in the arrival directions of
  ultra-high-energy cosmic rays from nearby radio galaxies.
\newblock \emph{The Astrophysical Journal}, 925\penalty0 (1):\penalty0 42, jan
  2022.
\newblock \doi{10.3847/1538-4357/ac3753}.
\newblock URL \url{https://doi.org/10.3847/1538-4357/ac3753}.

\bibitem[{The Telescope Array Collaboration}(2014)]{Abbasi_2014}
{The Telescope Array Collaboration}.
\newblock {I}ndications of intermediate-scale anisotropy of cosmic rays with
  energy greater than 57 {EeV} in the northern sky measured with the surface
  detector of the {T}elescope {A}rray {E}xperiment.
\newblock \emph{The Astrophysical Journal}, 790\penalty0 (2):\penalty0 L21, jul
  2014.
\newblock \doi{10.1088/2041-8205/790/2/l21}.
\newblock URL \url{https://doi.org/10.1088/2041-8205/790/2/l21}.

\bibitem[Romero et~al.(1996)Romero, Combi, {Perez Bergliaffa}, and
  Anchordoqui]{ROMERO1996279}
Gustavo~E. Romero, Jorge~A. Combi, Santiago~E. {Perez Bergliaffa}, and Luis~A.
  Anchordoqui.
\newblock Centaurus {A} as a source of extragalactic cosmic rays with arrival
  energies well beyond the {GZK} cutoff.
\newblock \emph{Astroparticle Physics}, 5\penalty0 (3):\penalty0 279--283,
  1996.
\newblock ISSN 0927-6505.
\newblock \doi{https://doi.org/10.1016/0927-6505(96)00029-1}.
\newblock URL
  \url{https://www.sciencedirect.com/science/article/pii/0927650596000291}.

\bibitem[Biermann and De~Souza(2012)]{biermann2012centaurus}
Peter~L Biermann and Vitor De~Souza.
\newblock {C}entaurus {A}: {T}he extragalactic source of cosmic rays with
  energies above the knee.
\newblock \emph{The Astrophysical Journal}, 746\penalty0 (1):\penalty0 72,
  2012.

\bibitem[Liu et~al.(2012)Liu, Wang, Wang, and Taylor]{liu2012excess}
Ruo-Yu Liu, Xiang-Yu Wang, Wei Wang, and Andrew~M Taylor.
\newblock On the excess of ultra-high energy cosmic rays in the direction of
  {C}entaurus {A}.
\newblock \emph{The Astrophysical Journal}, 755\penalty0 (2):\penalty0 139,
  2012.

\bibitem[Wykes et~al.(2013)Wykes, Croston, Hardcastle, Eilek, Biermann,
  Achterberg, Bray, Lazarian, Haverkorn, Protheroe, et~al.]{wykes2013mass}
Sarka Wykes, Judith~H Croston, Martin~J Hardcastle, Jean~A Eilek, Peter~L
  Biermann, Abraham Achterberg, Justin~D Bray, Alex Lazarian, Marijke
  Haverkorn, Ray~J Protheroe, et~al.
\newblock Mass entrainment and turbulence-driven acceleration of ultra-high
  energy cosmic rays in {C}entaurus {A}.
\newblock \emph{Astronomy \& Astrophysics}, 558:\penalty0 A19, 2013.

\bibitem[Farrar et~al.(2013)Farrar, Jansson, Feain, and Gaensler]{Farrar_2013}
Glennys~R Farrar, Ronnie Jansson, Ilana~J Feain, and B.M Gaensler.
\newblock Galactic magnetic deflections and {Centaurus A} as a {UHECR} source.
\newblock \emph{Journal of Cosmology and Astroparticle Physics}, 2013\penalty0
  (01):\penalty0 023--023, jan 2013.
\newblock \doi{10.1088/1475-7516/2013/01/023}.
\newblock URL \url{https://doi.org/10.1088/1475-7516/2013/01/023}.

\bibitem[Wykes et~al.(2018)Wykes, Taylor, Bray, Hardcastle, and
  Hillas]{wykes2018uhecr}
Sarka Wykes, Andrew~M Taylor, Justin~D Bray, Martin~J Hardcastle, and Michael
  Hillas.
\newblock {UHECR} propagation from {C}entaurus {A}.
\newblock \emph{Nuclear and particle physics proceedings}, 297:\penalty0
  234--241, 2018.

\bibitem[de~Oliveira and de~Souza(2021)]{deoliveira2021probing}
Cain{\~a} de~Oliveira and Vitor de~Souza.
\newblock Probing {UHECR} production in {Centaurus A} using secondary neutrinos
  and gamma-rays.
\newblock \emph{The European Physical Journal C}, 81\penalty0 (6):\penalty0
  1--17, 2021.

\bibitem[Caccianiga(2019)]{caccianiga2019anisotropies}
Lorenzo Caccianiga.
\newblock Anisotropies of the highest energy cosmic-ray events recorded by the
  {P}ierre {A}uger {O}bservatory in 15 years of operation.
\newblock In \emph{36th International Cosmic Ray Conference}, volume 358, page
  206. SISSA Medialab, 2019.

\bibitem[Dolag et~al.(2009)Dolag, Kachelrie{\ss}, and Semikoz]{Dolag_2009}
K~Dolag, M~Kachelrie{\ss}, and D.V Semikoz.
\newblock {UHECR} observations and lensing in the magnetic field of the {Virgo}
  cluster.
\newblock \emph{Journal of Cosmology and Astroparticle Physics}, 2009\penalty0
  (01):\penalty0 033--033, jan 2009.
\newblock \doi{10.1088/1475-7516/2009/01/033}.
\newblock URL \url{https://doi.org/10.1088/1475-7516/2009/01/033}.

\bibitem[Abbasi et~al.(2018)Abbasi, Abe, Abu-Zayyad, Allen, Azuma, Barcikowski,
  Belz, Bergman, Blake, Cady, Cheon, Chiba, Chikawa, di~Matteo, Fujii, Fujita,
  Fukushima, Furlich, Goto, Hanlon, Hayashi, Hayashi, Hayashida, Hibino, Honda,
  Ikeda, Inoue, Ishii, Ishimori, Ito, Ivanov, Jeong, Jeong, Jui, Kadota,
  Kakimoto, Kalashev, Kasahara, Kawai, Kawakami, Kawana, Kawata, Kido, Kim,
  Kim, Kim, Kishigami, Kitamura, Kitamura, Kuzmin, Kuznetsov, Kwon, Lee,
  Lubsandorzhiev, Lundquist, Machida, Martens, Matsuyama, Matthews, Mayta,
  Minamino, Mukai, Myers, Nagasawa, Nagataki, Nakamura, Nakamura, Nonaka, Oda,
  Ogio, Ogura, Ohnishi, Ohoka, Okuda, Omura, Ono, Onogi, Oshima, Ozawa, Park,
  Pshirkov, Remington, Rodriguez, Rubtsov, Ryu, Sagawa, Sahara, Saito, Saito,
  Sakaki, Sakurai, Scott, Seki, Sekino, Shah, Shibata, Shibata, Shimodaira,
  Shin, Shin, Smith, Sokolsky, Stokes, Stratton, Stroman, Suzawa, Takagi,
  Takahashi, Takamura, Takeda, Takeishi, Taketa, Takita, Tameda, Tanaka,
  Tanaka, Tanaka, Thomas, Thomson, Tinyakov, Tkachev, Tokuno, Tomida, Troitsky,
  Tsunesada, Tsutsumi, Uchihori, Udo, Urban, Wong, Yamamoto, Yamane, Yamaoka,
  Yamazaki, Yang, Yashiro, Yoneda, Yoshida, Yoshii, Zhezher, and
  Zundel]{Abbasi_2018}
R.~U. Abbasi, M.~Abe, T.~Abu-Zayyad, M.~Allen, R.~Azuma, E.~Barcikowski, J.~W.
  Belz, D.~R. Bergman, S.~A. Blake, R.~Cady, B.~G. Cheon, J.~Chiba, M.~Chikawa,
  A.~di~Matteo, T.~Fujii, K.~Fujita, M.~Fukushima, G.~Furlich, T.~Goto,
  W.~Hanlon, M.~Hayashi, Y.~Hayashi, N.~Hayashida, K.~Hibino, K.~Honda,
  D.~Ikeda, N.~Inoue, T.~Ishii, R.~Ishimori, H.~Ito, D.~Ivanov, H.~M. Jeong,
  S.~Jeong, C.~C.~H. Jui, K.~Kadota, F.~Kakimoto, O.~Kalashev, K.~Kasahara,
  H.~Kawai, S.~Kawakami, S.~Kawana, K.~Kawata, E.~Kido, H.~B. Kim, J.~H. Kim,
  J.~H. Kim, S.~Kishigami, S.~Kitamura, Y.~Kitamura, V.~Kuzmin, M.~Kuznetsov,
  Y.~J. Kwon, K.~H. Lee, B.~Lubsandorzhiev, J.~P. Lundquist, K.~Machida,
  K.~Martens, T.~Matsuyama, J.~N. Matthews, R.~Mayta, M.~Minamino, K.~Mukai,
  I.~Myers, K.~Nagasawa, S.~Nagataki, R.~Nakamura, T.~Nakamura, T.~Nonaka,
  H.~Oda, S.~Ogio, J.~Ogura, M.~Ohnishi, H.~Ohoka, T.~Okuda, Y.~Omura, M.~Ono,
  R.~Onogi, A.~Oshima, S.~Ozawa, I.~H. Park, M.~S. Pshirkov, J.~Remington,
  D.~C. Rodriguez, G.~Rubtsov, D.~Ryu, H.~Sagawa, R.~Sahara, K.~Saito,
  Y.~Saito, N.~Sakaki, N.~Sakurai, L.~M. Scott, T.~Seki, K.~Sekino, P.~D. Shah,
  F.~Shibata, T.~Shibata, H.~Shimodaira, B.~K. Shin, H.~S. Shin, J.~D. Smith,
  P.~Sokolsky, B.~T. Stokes, S.~R. Stratton, T.~A. Stroman, T.~Suzawa,
  Y.~Takagi, Y.~Takahashi, M.~Takamura, M.~Takeda, R.~Takeishi, A.~Taketa,
  M.~Takita, Y.~Tameda, H.~Tanaka, K.~Tanaka, M.~Tanaka, S.~B. Thomas, G.~B.
  Thomson, P.~Tinyakov, I.~Tkachev, H.~Tokuno, T.~Tomida, S.~Troitsky,
  Y.~Tsunesada, K.~Tsutsumi, Y.~Uchihori, S.~Udo, F.~Urban, T.~Wong,
  M.~Yamamoto, R.~Yamane, H.~Yamaoka, K.~Yamazaki, J.~Yang, K.~Yashiro,
  Y.~Yoneda, S.~Yoshida, H.~Yoshii, Y.~Zhezher, and Z.~Zundel.
\newblock Testing a reported correlation between arrival directions of
  ultra-high-energy cosmic rays and a flux pattern from nearby starburst
  galaxies using telescope array data.
\newblock \emph{The Astrophysical Journal}, 867\penalty0 (2):\penalty0 L27, nov
  2018.
\newblock \doi{10.3847/2041-8213/aaebf9}.
\newblock URL \url{https://doi.org/10.3847/2041-8213/aaebf9}.

\bibitem[Erdmann et~al.(2016)Erdmann, Müller, Urban, and Wirtz]{ERDMANN201654}
M.~Erdmann, G.~Müller, M.~Urban, and M.~Wirtz.
\newblock The nuclear window to the extragalactic universe.
\newblock \emph{Astroparticle Physics}, 85:\penalty0 54--64, 2016.
\newblock ISSN 0927-6505.
\newblock \doi{https://doi.org/10.1016/j.astropartphys.2016.10.002}.
\newblock URL
  \url{https://www.sciencedirect.com/science/article/pii/S0927650516301451}.

\bibitem[Farrar and Sutherland(2019)]{Farrar_2019}
Glennys~R. Farrar and Michael~S. Sutherland.
\newblock Deflections of {UHECRs} in the galactic magnetic field.
\newblock \emph{Journal of Cosmology and Astroparticle Physics}, 2019\penalty0
  (05):\penalty0 004--004, may 2019.
\newblock \doi{10.1088/1475-7516/2019/05/004}.
\newblock URL \url{https://doi.org/10.1088/1475-7516/2019/05/004}.

\bibitem[Sigl et~al.(2003{\natexlab{a}})Sigl, Miniati, and
  Ensslin]{https://doi.org/10.48550/arxiv.astro-ph/0309695}
Guenter Sigl, Francesco Miniati, and Torsten Ensslin.
\newblock Signatures of magnetized large scale structure in ultra-high energy
  cosmic rays, 2003{\natexlab{a}}.
\newblock URL \url{https://arxiv.org/abs/astro-ph/0309695}.

\bibitem[Sigl et~al.(2003{\natexlab{b}})Sigl, Miniati, and
  Ensslin]{PhysRevD.68.043002}
G\"unter Sigl, Francesco Miniati, and Torsten~A. Ensslin.
\newblock Ultrahigh energy cosmic rays in a structured and magnetized universe.
\newblock \emph{Phys. Rev. D}, 68:\penalty0 043002, Aug 2003{\natexlab{b}}.
\newblock \doi{10.1103/PhysRevD.68.043002}.
\newblock URL \url{https://link.aps.org/doi/10.1103/PhysRevD.68.043002}.

\bibitem[Dolag et~al.(2005)Dolag, Grasso, Springel, and Tkachev]{Dolag_2005}
Klaus Dolag, Dario Grasso, Volker Springel, and Igor Tkachev.
\newblock Constrained simulations of the magnetic field in the local universe
  and the propagation of ultrahigh energy cosmic rays.
\newblock \emph{Journal of Cosmology and Astroparticle Physics}, 2005\penalty0
  (01):\penalty0 009--009, jan 2005.
\newblock \doi{10.1088/1475-7516/2005/01/009}.
\newblock URL \url{https://doi.org/10.1088/1475-7516/2005/01/009}.

\bibitem[Alves~Batista et~al.(2017)Alves~Batista, Shin, Devriendt, Semikoz, and
  Sigl]{PhysRevD.96.023010}
Rafael Alves~Batista, Min-Su Shin, Julien Devriendt, Dmitri Semikoz, and
  Guenter Sigl.
\newblock Implications of strong intergalactic magnetic fields for
  ultrahigh-energy cosmic-ray astronomy.
\newblock \emph{Phys. Rev. D}, 96:\penalty0 023010, Jul 2017.
\newblock \doi{10.1103/PhysRevD.96.023010}.
\newblock URL \url{https://link.aps.org/doi/10.1103/PhysRevD.96.023010}.

\bibitem[Subramanian(2016)]{Subramanian_2016}
Kandaswamy Subramanian.
\newblock The origin, evolution and signatures of primordial magnetic fields.
\newblock \emph{Reports on Progress in Physics}, 79\penalty0 (7):\penalty0
  076901, may 2016.
\newblock \doi{10.1088/0034-4885/79/7/076901}.
\newblock URL \url{https://doi.org/10.1088/0034-4885/79/7/076901}.

\bibitem[Hackstein et~al.(2018)Hackstein, Vazza, Br{\"u}ggen, Sorce, and
  Gottl{\"o}ber]{hackstein2018simulations}
Stefan Hackstein, Franco Vazza, Marcus Br{\"u}ggen, Jenny~G Sorce, and Stefan
  Gottl{\"o}ber.
\newblock Simulations of ultra-high energy cosmic rays in the local universe
  and the origin of cosmic magnetic fields.
\newblock \emph{Monthly Notices of the Royal Astronomical Society},
  475\penalty0 (2):\penalty0 2519--2529, 2018.

\bibitem[{Sun, X. H.} et~al.(2008){Sun, X. H.}, {Reich, W.}, {Waelkens, A.},
  and {En\ss{}lin, T. A.}]{sun_2008_gmf}
{Sun, X. H.}, {Reich, W.}, {Waelkens, A.}, and {En\ss{}lin, T. A.}
\newblock Radio observational constraints on galactic 3d-emission models.
\newblock \emph{A\&A}, 477\penalty0 (2):\penalty0 573--592, 2008.
\newblock \doi{10.1051/0004-6361:20078671}.
\newblock URL \url{https://doi.org/10.1051/0004-6361:20078671}.

\bibitem[Jansson and Farrar(2012{\natexlab{a}})]{Jansson_2012a}
Ronnie Jansson and Glennys~R. Farrar.
\newblock A new model of the {G}alactic magnetic field.
\newblock \emph{The Astrophysical Journal}, 757\penalty0 (1):\penalty0 14, Aug
  2012{\natexlab{a}}.
\newblock ISSN 1538-4357.
\newblock \doi{10.1088/0004-637x/757/1/14}.
\newblock URL \url{http://dx.doi.org/10.1088/0004-637X/757/1/14}.

\bibitem[Jansson and Farrar(2012{\natexlab{b}})]{Jansson_2012b}
Ronnie Jansson and Glennys~R. Farrar.
\newblock {T}he {G}alactic {M}agnetic {F}ield.
\newblock \emph{The Astrophysical Journal}, 761\penalty0 (1):\penalty0 L11, nov
  2012{\natexlab{b}}.
\newblock \doi{10.1088/2041-8205/761/1/l11}.
\newblock URL \url{https://doi.org/10.1088/2041-8205/761/1/l11}.

\bibitem[Pshirkov et~al.(2011)Pshirkov, Tinyakov, Kronberg, and
  Newton-McGee]{Pshirkov_2011_gmf}
M.~S. Pshirkov, P.~G. Tinyakov, P.~P. Kronberg, and K.~J. Newton-McGee.
\newblock {DERIVING} {THE} {GLOBAL} {STRUCTURE} {OF} {THE} {GALACTIC}
  {MAGNETIC} {FIELD} {FROM} {FARADAY} {ROTATION} {MEASURES} {OF}
  {EXTRAGALACTIC} {SOURCES}.
\newblock \emph{The Astrophysical Journal}, 738\penalty0 (2):\penalty0 192, aug
  2011.
\newblock \doi{10.1088/0004-637x/738/2/192}.
\newblock URL \url{https://doi.org/10.1088/0004-637x/738/2/192}.

\bibitem[Zatsepin and Kuz'min(1966)]{bib:zk}
Georgi~T Zatsepin and Vadem~A Kuz'min.
\newblock Upper limit of the spectrum of cosmic rays.
\newblock \emph{Soviet Journal of Experimental and Theoretical Physics
  Letters}, 4:\penalty0 78, 1966.

\bibitem[Greisen(1966)]{Greisen1966}
Kenneth Greisen.
\newblock End to the cosmic-ray spectrum?
\newblock \emph{Phys. Rev. Lett.}, 16\penalty0 (17):\penalty0 748--, April
  1966.
\newblock URL \url{http://link.aps.org/abstract/PRL/v16/p748}.

\bibitem[Deligny et~al.(2004)Deligny, Letessier-Selvon, and
  Parizot]{DELIGNY2004609}
Olivier Deligny, Antoine Letessier-Selvon, and Etienne Parizot.
\newblock Magnetic horizons of uhecr sources and the gzk feature.
\newblock \emph{Astroparticle Physics}, 21\penalty0 (6):\penalty0 609--615,
  2004.
\newblock ISSN 0927-6505.
\newblock \doi{https://doi.org/10.1016/j.astropartphys.2004.04.012}.
\newblock URL
  \url{https://www.sciencedirect.com/science/article/pii/S0927650504000908}.

\bibitem[Taylor et~al.(2011)Taylor, Ahlers, and Aharonian]{taylor2011need}
Andrew~M Taylor, Markus Ahlers, and Felix~A Aharonian.
\newblock Need for a local source of ultrahigh-energy cosmic-ray nuclei.
\newblock \emph{Physical Review D}, 84\penalty0 (10):\penalty0 105007, 2011.

\bibitem[Alves~Batista and Sigl(2014)]{AlvesBatista:2014tdg}
Rafael Alves~Batista and G\"unter Sigl.
\newblock {Magnetic horizons of ultra-high energy cosmic rays}.
\newblock In \emph{{20th International Conference on Particles and Nuclei}},
  pages 387--390, 9 2014.
\newblock \doi{10.3204/DESY-PROC-2014-04/127}.

\bibitem[Lang et~al.(2020)Lang, Taylor, Ahlers, and
  de~Souza]{lang2020revisiting}
Rodrigo~Guedes Lang, Andrew~M Taylor, Markus Ahlers, and Vitor de~Souza.
\newblock Revisiting the distance to the nearest ultrahigh energy cosmic ray
  source: Effects of extragalactic magnetic fields.
\newblock \emph{Physical Review D}, 102\penalty0 (6):\penalty0 063012, 2020.

\bibitem[Lang et~al.(2021)Lang, Taylor, and de~Souza]{lang2020_dipole}
Rodrigo~Guedes Lang, Andrew~M. Taylor, and Vitor de~Souza.
\newblock Ultrahigh-energy cosmic rays dipole and beyond.
\newblock \emph{Phys. Rev. D}, 103:\penalty0 063005, Mar 2021.
\newblock \doi{10.1103/PhysRevD.103.063005}.
\newblock URL \url{https://link.aps.org/doi/10.1103/PhysRevD.103.063005}.

\bibitem[Cronin(1999)]{RevModPhys.71.S165}
James~W. Cronin.
\newblock Cosmic rays: the most energetic particles in the universe.
\newblock \emph{Rev. Mod. Phys.}, 71:\penalty0 S165--S172, Mar 1999.
\newblock \doi{10.1103/RevModPhys.71.S165}.
\newblock URL \url{https://link.aps.org/doi/10.1103/RevModPhys.71.S165}.

\bibitem[Supanitsky and de~Souza(2013)]{supanitsky2013upper}
Alberto~Daniel Supanitsky and V~de~Souza.
\newblock An upper limit on the cosmic-ray luminosity of individual sources
  from gamma-ray observations.
\newblock \emph{Journal of Cosmology and Astroparticle Physics}, 2013\penalty0
  (12):\penalty0 023, 2013.

\bibitem[Kotera and Olinto(2011)]{kotera2011astrophysics}
Kumiko Kotera and Angela~V Olinto.
\newblock The astrophysics of ultrahigh-energy cosmic rays.
\newblock \emph{Annual Review of Astronomy and Astrophysics}, 49:\penalty0
  119--153, 2011.

\bibitem[Aab et~al.(2020{\natexlab{b}})Aab, Abreu, Aglietta, Albury, Allekotte,
  Almela, Alvarez~Castillo, Alvarez-Mu\~niz, Alves~Batista, Anastasi,
  Anchordoqui, Andrada, Andringa, Aramo, Ara\'ujo~Ferreira, Asorey, Assis,
  Avila, Badescu, Bakalova, Balaceanu, Barbato, Barreira~Luz, Becker, Bellido,
  Berat, Bertaina, Bertou, Biermann, Bister, Biteau, Blanco, Blazek, Bleve,
  Boh\'a\ifmmode~\check{c}\else \v{c}\fi{}ov\'a, Boncioli, Bonifazi,
  Bonneau~Arbeletche, Borodai, Botti, Brack, Bretz, Briechle, Buchholz, Bueno,
  Buitink, Buscemi, Caballero-Mora, Caccianiga, Calcagni, Cancio, Canfora,
  Caracas, Carceller, Caruso, Castellina, Catalani, Cataldi, Cazon, Cerda,
  Chinellato, Choi, Chudoba, Chytka, Clay, Cobos~Cerutti, Colalillo, Coleman,
  Coluccia, Concei\ifmmode \mbox{\c{c}}\else~\c{c}\fi{}\ ao, Condorelli,
  Consolati, Contreras, Convenga, Covault, Dasso, Daumiller, Dawson, Day,
  de~Almeida, de~Jes\'us, de~Jong, De~Mauro, de~Mello~Neto, De~Mitri,
  de~Oliveira, de~Oliveira~Franco, de~Souza, De~Vito, Debatin, del R\'{\i}o,
  Deligny, Dembinski, Dhital, Di~Giulio, Di~Matteo, D\'{\i}az~Castro,
  Dobrigkeit, D'Olivo, Dorosti, dos Anjos, Dova, Ebr, Engel, Epicoco, Erdmann,
  Escobar, Etchegoyen, Falcke, Farmer, Farrar, Fauth, Fazzini, Feldbusch, Fenu,
  Fick, Figueira, Filip\ifmmode \check{c}\else
  \v{c}\fi{}i\ifmmode~\check{c}\else \v{c}\fi{}, Fodran, Freire, Fujii, Fuster,
  Galea, Galelli, Garc\'{\i}a, Garcia~Vegas, Gemmeke, Gesualdi, Gherghel-Lascu,
  Ghia, Giaccari, Giammarchi, Giller, Glombitza, Gobbi, Gollan, Golup,
  G\'omez~Berisso, G\'omez~Vitale, Gongora, Gonz\'alez, Goos, G\'ora, Gorgi,
  Gottowik, Grubb, Guarino, Guedes, Guido, Hahn, Halliday, Hampel, Hansen,
  Harari, Harvey, Haungs, Hebbeker, Heck, Hill, Hojvat, H\"orandel, Horvath,
  Hrabovsk\'y, Huege, Hulsman, Insolia, Isar, Johnsen, Jurysek, K\"a\"ap\"a,
  Kampert, Keilhauer, Kemp, Klages, Kleifges, Kleinfeller, K\"opke,
  Kukec~Mezek, Lago, LaHurd, Lang, Leigui~de Oliveira, Lenok, Letessier-Selvon,
  Lhenry-Yvon, Lo~Presti, Lopes, L\'opez, Lorek, Luce, Lucero, Machado~Payeras,
  Malacari, Mancarella, Mandat, Manning, Manshanden, Mantsch, Marafico,
  Mariazzi, Mari\ifmmode~\mbox{\c{s}}\else \c{s}\fi{}, Marsella, Martello,
  Martinez, Mart\'{\i}nez~Bravo, Mastrodicasa, Mathes, Matthews, Matthiae,
  Mayotte, Mazur, Medina-Tanco, Melo, Menshikov, Merenda, Michal, Micheletti,
  Miramonti, Mockler, Mollerach, Montanet, Morello, Mostaf\'a, M\"uller,
  Muller, Mulrey, Mussa, Muzio, Namasaka, Nellen, Nguyen, Niculescu-Oglinzanu,
  Niechciol, Nitz, Nosek, Novotny, No\ifmmode~\check{z}\else \v{z}\fi{}ka,
  Nucita, N\'u\~nez, Palatka, Pallotta, Panetta, Papenbreer, Parente, Parra,
  Pech, Pedreira, P\ifmmode~\mbox{\c{e}}\else \c{e}\fi{}kala, Pelayo, Pe\~na
  Rodriguez, Perez~Armand, Perlin, Perrone, Peters, Petrera, Pierog, Pimenta,
  Pirronello, Platino, Pont, Pothast, Privitera, Prouza, Puyleart, Querchfeld,
  Rautenberg, Ravignani, Reininghaus, Ridky, Riehn, Risse, Ristori, Rizi,
  Rodrigues~de Carvalho, Rodriguez~Fernandez, Rodriguez~Rojo, Roncoroni, Roth,
  Roulet, Rovero, Ruehl, Saffi, Saftoiu, Salamida, Salazar, Salina,
  Sanabria~Gomez, S\'anchez, Santos, Santos, Sarazin, Sarmento, Sarmiento-Cano,
  Sato, Savina, Sch\"afer, Scherini, Schieler, Schimassek, Schimp, Schl\"uter,
  Schmidt, Scholten, Schov\'anek, Schr\"oder, Schr\"oder, Schulz, Sciutto,
  Scornavacche, Shellard, Sigl, Silli, Sima, \ifmmode~\check{S}\else
  \v{S}\fi{}m\'{\i}da, Sommers, Soriano, Souchard, Squartini, Stadelmaier,
  Stanca, Stani\ifmmode~\check{c}\else \v{c}\fi{}, Stasielak, Stassi, Streich,
  Su\'arez-Dur\'an, Sudholz, Suomij\"arvi, Supanitsky, \ifmmode~\check{S}\else
  \v{S}\fi{}up\'{\i}k, Szadkowski, Taboada, Tapia, Timmermans, Tkachenko,
  Tobiska, Todero~Peixoto, Tom\'e, Torralba~Elipe, Travaini, Travnicek,
  Trimarelli, Trini, Tueros, Ulrich, Unger, Urban, Vaclavek, Vacula,
  Vald\'es~Galicia, Vali\~no, Valore, van Vliet, Varela, Vargas~C\'ardenas,
  V\'asquez-Ram\'{\i}rez, Veberi\ifmmode~\check{c}\else \v{c}\fi{}, Ventura,
  Vergara~Quispe, Verzi, Vicha, Villase\~nor, Vink, Vorobiov, Wahlberg, Watson,
  Weber, Weindl, Wiencke, Wilczy\ifmmode~\acute{n}\else \'{n}\fi{}ski, Winchen,
  Wirtz, Wittkowski, Wundheiler, Yushkov, Zapparrata, Zas, Zavrtanik,
  Zavrtanik, Zehrer, Zepeda, Ziolkowski, and Zuccarello]{PhysRevD.102.062005}
A.~Aab, P.~Abreu, M.~Aglietta, J.~M. Albury, I.~Allekotte, A.~Almela,
  J.~Alvarez~Castillo, J.~Alvarez-Mu\~niz, R.~Alves~Batista, G.~A. Anastasi,
  L.~Anchordoqui, B.~Andrada, S.~Andringa, C.~Aramo, P.~R. Ara\'ujo~Ferreira,
  H.~Asorey, P.~Assis, G.~Avila, A.~M. Badescu, A.~Bakalova, A.~Balaceanu,
  F.~Barbato, R.~J. Barreira~Luz, K.~H. Becker, J.~A. Bellido, C.~Berat, M.~E.
  Bertaina, X.~Bertou, P.~L. Biermann, T.~Bister, J.~Biteau, A.~Blanco,
  J.~Blazek, C.~Bleve, M.~Boh\'a\ifmmode~\check{c}\else \v{c}\fi{}ov\'a,
  D.~Boncioli, C.~Bonifazi, L.~Bonneau~Arbeletche, N.~Borodai, A.~M. Botti,
  J.~Brack, T.~Bretz, F.~L. Briechle, P.~Buchholz, A.~Bueno, S.~Buitink,
  M.~Buscemi, K.~S. Caballero-Mora, L.~Caccianiga, L.~Calcagni, A.~Cancio,
  F.~Canfora, I.~Caracas, J.~M. Carceller, R.~Caruso, A.~Castellina,
  F.~Catalani, G.~Cataldi, L.~Cazon, M.~Cerda, J.~A. Chinellato, K.~Choi,
  J.~Chudoba, L.~Chytka, R.~W. Clay, A.~C. Cobos~Cerutti, R.~Colalillo,
  A.~Coleman, M.~R. Coluccia, R.~Concei\ifmmode \mbox{\c{c}}\else~\c{c}\fi{}\
  ao, A.~Condorelli, G.~Consolati, F.~Contreras, F.~Convenga, C.~E. Covault,
  S.~Dasso, K.~Daumiller, B.~R. Dawson, J.~A. Day, R.~M. de~Almeida,
  J.~de~Jes\'us, S.~J. de~Jong, G.~De~Mauro, J.~R.~T. de~Mello~Neto,
  I.~De~Mitri, J.~de~Oliveira, D.~de~Oliveira~Franco, V.~de~Souza, E.~De~Vito,
  J.~Debatin, M.~del R\'{\i}o, O.~Deligny, H.~Dembinski, N.~Dhital,
  C.~Di~Giulio, A.~Di~Matteo, M.~L. D\'{\i}az~Castro, C.~Dobrigkeit, J.~C.
  D'Olivo, Q.~Dorosti, R.~C. dos Anjos, M.~T. Dova, J.~Ebr, R.~Engel,
  I.~Epicoco, M.~Erdmann, C.~O. Escobar, A.~Etchegoyen, H.~Falcke, J.~Farmer,
  G.~Farrar, A.~C. Fauth, N.~Fazzini, F.~Feldbusch, F.~Fenu, B.~Fick, J.~M.
  Figueira, A.~Filip\ifmmode \check{c}\else \v{c}\fi{}i\ifmmode~\check{c}\else
  \v{c}\fi{}, T.~Fodran, M.~M. Freire, T.~Fujii, A.~Fuster, C.~Galea,
  C.~Galelli, B.~Garc\'{\i}a, A.~L. Garcia~Vegas, H.~Gemmeke, F.~Gesualdi,
  A.~Gherghel-Lascu, P.~L. Ghia, U.~Giaccari, M.~Giammarchi, M.~Giller,
  J.~Glombitza, F.~Gobbi, F.~Gollan, G.~Golup, M.~G\'omez~Berisso, P.~F.
  G\'omez~Vitale, J.~P. Gongora, N.~Gonz\'alez, I.~Goos, D.~G\'ora, A.~Gorgi,
  M.~Gottowik, T.~D. Grubb, F.~Guarino, G.~P. Guedes, E.~Guido, S.~Hahn,
  R.~Halliday, M.~R. Hampel, P.~Hansen, D.~Harari, V.~M. Harvey, A.~Haungs,
  T.~Hebbeker, D.~Heck, G.~C. Hill, C.~Hojvat, J.~R. H\"orandel, P.~Horvath,
  M.~Hrabovsk\'y, T.~Huege, J.~Hulsman, A.~Insolia, P.~G. Isar, J.~A. Johnsen,
  J.~Jurysek, A.~K\"a\"ap\"a, K.~H. Kampert, B.~Keilhauer, J.~Kemp, H.~O.
  Klages, M.~Kleifges, J.~Kleinfeller, M.~K\"opke, G.~Kukec~Mezek, B.~L. Lago,
  D.~LaHurd, R.~G. Lang, M.~A. Leigui~de Oliveira, V.~Lenok,
  A.~Letessier-Selvon, I.~Lhenry-Yvon, D.~Lo~Presti, L.~Lopes, R.~L\'opez,
  R.~Lorek, Q.~Luce, A.~Lucero, A.~Machado~Payeras, M.~Malacari, G.~Mancarella,
  D.~Mandat, B.~C. Manning, J.~Manshanden, P.~Mantsch, S.~Marafico, A.~G.
  Mariazzi, I.~C. Mari\ifmmode~\mbox{\c{s}}\else \c{s}\fi{}, G.~Marsella,
  D.~Martello, H.~Martinez, O.~Mart\'{\i}nez~Bravo, M.~Mastrodicasa, H.~J.
  Mathes, J.~Matthews, G.~Matthiae, E.~Mayotte, P.~O. Mazur, G.~Medina-Tanco,
  D.~Melo, A.~Menshikov, K.-D. Merenda, S.~Michal, M.~I. Micheletti,
  L.~Miramonti, D.~Mockler, S.~Mollerach, F.~Montanet, C.~Morello,
  M.~Mostaf\'a, A.~L. M\"uller, M.~A. Muller, K.~Mulrey, R.~Mussa, M.~Muzio,
  W.~M. Namasaka, L.~Nellen, P.~H. Nguyen, M.~Niculescu-Oglinzanu,
  M.~Niechciol, D.~Nitz, D.~Nosek, V.~Novotny, L.~No\ifmmode~\check{z}\else
  \v{z}\fi{}ka, A.~Nucita, L.~A. N\'u\~nez, M.~Palatka, J.~Pallotta, M.~P.
  Panetta, P.~Papenbreer, G.~Parente, A.~Parra, M.~Pech, F.~Pedreira,
  J.~P\ifmmode~\mbox{\c{e}}\else \c{e}\fi{}kala, R.~Pelayo, J.~Pe\~na
  Rodriguez, J.~Perez~Armand, M.~Perlin, L.~Perrone, C.~Peters, S.~Petrera,
  T.~Pierog, M.~Pimenta, V.~Pirronello, M.~Platino, B.~Pont, M.~Pothast,
  P.~Privitera, M.~Prouza, A.~Puyleart, S.~Querchfeld, J.~Rautenberg,
  D.~Ravignani, M.~Reininghaus, J.~Ridky, F.~Riehn, M.~Risse, P.~Ristori,
  V.~Rizi, W.~Rodrigues~de Carvalho, G.~Rodriguez~Fernandez, J.~Rodriguez~Rojo,
  M.~J. Roncoroni, M.~Roth, E.~Roulet, A.~C. Rovero, P.~Ruehl, S.~J. Saffi,
  A.~Saftoiu, F.~Salamida, H.~Salazar, G.~Salina, J.~D. Sanabria~Gomez,
  F.~S\'anchez, E.~M. Santos, E.~Santos, F.~Sarazin, R.~Sarmento,
  C.~Sarmiento-Cano, R.~Sato, P.~Savina, C.~Sch\"afer, V.~Scherini,
  H.~Schieler, M.~Schimassek, M.~Schimp, F.~Schl\"uter, D.~Schmidt,
  O.~Scholten, P.~Schov\'anek, F.~G. Schr\"oder, S.~Schr\"oder, A.~Schulz,
  S.~J. Sciutto, M.~Scornavacche, R.~C. Shellard, G.~Sigl, G.~Silli, O.~Sima,
  R.~\ifmmode~\check{S}\else \v{S}\fi{}m\'{\i}da, P.~Sommers, J.~F. Soriano,
  J.~Souchard, R.~Squartini, M.~Stadelmaier, D.~Stanca,
  S.~Stani\ifmmode~\check{c}\else \v{c}\fi{}, J.~Stasielak, P.~Stassi,
  A.~Streich, M.~Su\'arez-Dur\'an, T.~Sudholz, T.~Suomij\"arvi, A.~D.
  Supanitsky, J.~\ifmmode~\check{S}\else \v{S}\fi{}up\'{\i}k, Z.~Szadkowski,
  A.~Taboada, A.~Tapia, C.~Timmermans, O.~Tkachenko, P.~Tobiska, C.~J.
  Todero~Peixoto, B.~Tom\'e, G.~Torralba~Elipe, A.~Travaini, P.~Travnicek,
  C.~Trimarelli, M.~Trini, M.~Tueros, R.~Ulrich, M.~Unger, M.~Urban,
  L.~Vaclavek, M.~Vacula, J.~F. Vald\'es~Galicia, I.~Vali\~no, L.~Valore,
  A.~van Vliet, E.~Varela, B.~Vargas~C\'ardenas, A.~V\'asquez-Ram\'{\i}rez,
  D.~Veberi\ifmmode~\check{c}\else \v{c}\fi{}, C.~Ventura, I.~D.
  Vergara~Quispe, V.~Verzi, J.~Vicha, L.~Villase\~nor, J.~Vink, S.~Vorobiov,
  H.~Wahlberg, A.~A. Watson, M.~Weber, A.~Weindl, L.~Wiencke,
  H.~Wilczy\ifmmode~\acute{n}\else \'{n}\fi{}ski, T.~Winchen, M.~Wirtz,
  D.~Wittkowski, B.~Wundheiler, A.~Yushkov, O.~Zapparrata, E.~Zas,
  D.~Zavrtanik, M.~Zavrtanik, L.~Zehrer, A.~Zepeda, M.~Ziolkowski, and
  F.~Zuccarello.
\newblock Measurement of the cosmic-ray energy spectrum above
  $2.5\times10^{18}$~ev using the pierre auger observatory.
\newblock \emph{Phys. Rev. D}, 102:\penalty0 062005, Sep 2020{\natexlab{b}}.
\newblock \doi{10.1103/PhysRevD.102.062005}.
\newblock URL \url{https://link.aps.org/doi/10.1103/PhysRevD.102.062005}.

\bibitem[Ackermann et~al.(2012)Ackermann, Ajello, Allafort, Baldini, Ballet,
  Bastieri, Bechtol, Bellazzini, Berenji, Bloom, Bonamente, Borgland, Bouvier,
  Bregeon, Brigida, Bruel, Buehler, Buson, Caliandro, Cameron, Caraveo,
  Casandjian, Cecchi, Charles, Chekhtman, Cheung, Chiang, Cillis, Ciprini,
  Claus, Cohen-Tanugi, Conrad, Cutini, de~Palma, Dermer, Digel, do~Couto~e
  Silva, Drell, Drlica-Wagner, Favuzzi, Fegan, Fortin, Fukazawa, Funk, Fusco,
  Gargano, Gasparrini, Germani, Giglietto, Giordano, Glanzman, Godfrey,
  Grenier, Guiriec, Gustafsson, Hadasch, Hayashida, Hays, Hughes,
  J{\'{o}}hannesson, Johnson, Kamae, Katagiri, Kataoka, Knödlseder, Kuss,
  Lande, Longo, Loparco, Lott, Lovellette, Lubrano, Madejski, Martin,
  Mazziotta, McEnery, Michelson, Mizuno, Monte, Monzani, Morselli, Moskalenko,
  Murgia, Nishino, Norris, Nuss, Ohno, Ohsugi, Okumura, Omodei, Orlando, Ozaki,
  Parent, Persic, Pesce-Rollins, Petrosian, Pierbattista, Piron, Pivato,
  Porter, Rain{\`{o}}, Rando, Razzano, Reimer, Reimer, Ritz, Roth, Sbarra,
  Sgr{\`{o}}, Siskind, Spandre, Spinelli, Stawarz, Strong, Takahashi, Tanaka,
  Thayer, Tibaldo, Tinivella, Torres, Tosti, Troja, Uchiyama, Vandenbroucke,
  Vianello, Vitale, Waite, Wood, and Yang]{Ackermann_2012}
M.~Ackermann, M.~Ajello, A.~Allafort, L.~Baldini, J.~Ballet, D.~Bastieri,
  K.~Bechtol, R.~Bellazzini, B.~Berenji, E.~D. Bloom, E.~Bonamente, A.~W.
  Borgland, A.~Bouvier, J.~Bregeon, M.~Brigida, P.~Bruel, R.~Buehler, S.~Buson,
  G.~A. Caliandro, R.~A. Cameron, P.~A. Caraveo, J.~M. Casandjian, C.~Cecchi,
  E.~Charles, A.~Chekhtman, C.~C. Cheung, J.~Chiang, A.~N. Cillis, S.~Ciprini,
  R.~Claus, J.~Cohen-Tanugi, J.~Conrad, S.~Cutini, F.~de~Palma, C.~D. Dermer,
  S.~W. Digel, E.~do~Couto~e Silva, P.~S. Drell, A.~Drlica-Wagner, C.~Favuzzi,
  S.~J. Fegan, P.~Fortin, Y.~Fukazawa, S.~Funk, P.~Fusco, F.~Gargano,
  D.~Gasparrini, S.~Germani, N.~Giglietto, F.~Giordano, T.~Glanzman,
  G.~Godfrey, I.~A. Grenier, S.~Guiriec, M.~Gustafsson, D.~Hadasch,
  M.~Hayashida, E.~Hays, R.~E. Hughes, G.~J{\'{o}}hannesson, A.~S. Johnson,
  T.~Kamae, H.~Katagiri, J.~Kataoka, J.~Knödlseder, M.~Kuss, J.~Lande,
  F.~Longo, F.~Loparco, B.~Lott, M.~N. Lovellette, P.~Lubrano, G.~M. Madejski,
  P.~Martin, M.~N. Mazziotta, J.~E. McEnery, P.~F. Michelson, T.~Mizuno,
  C.~Monte, M.~E. Monzani, A.~Morselli, I.~V. Moskalenko, S.~Murgia,
  S.~Nishino, J.~P. Norris, E.~Nuss, M.~Ohno, T.~Ohsugi, A.~Okumura, N.~Omodei,
  E.~Orlando, M.~Ozaki, D.~Parent, M.~Persic, M.~Pesce-Rollins, V.~Petrosian,
  M.~Pierbattista, F.~Piron, G.~Pivato, T.~A. Porter, S.~Rain{\`{o}}, R.~Rando,
  M.~Razzano, A.~Reimer, O.~Reimer, S.~Ritz, M.~Roth, C.~Sbarra, C.~Sgr{\`{o}},
  E.~J. Siskind, G.~Spandre, P.~Spinelli, {\L}ukasz Stawarz, A.~W. Strong,
  H.~Takahashi, T.~Tanaka, J.~B. Thayer, L.~Tibaldo, M.~Tinivella, D.~F.
  Torres, G.~Tosti, E.~Troja, Y.~Uchiyama, J.~Vandenbroucke, G.~Vianello,
  V.~Vitale, A.~P. Waite, M.~Wood, and Z.~Yang.
\newblock {GeV} {OBSERVATIONS} {OF} {STAR}-{FORMING} {GALAXIES} {WITH} {THE}
  \textit{{FERMI}} {LARGE} {AREA} {TELESCOPE}.
\newblock \emph{The Astrophysical Journal}, 755\penalty0 (2):\penalty0 164, aug
  2012.
\newblock \doi{10.1088/0004-637x/755/2/164}.
\newblock URL \url{https://doi.org/10.1088/0004-637x/755/2/164}.

\bibitem[{Sahakyan, N.} et~al.(2018){Sahakyan, N.}, {Baghmanyan, V.}, and
  {Zargaryan, D.}]{sahakyan2018}
{Sahakyan, N.}, {Baghmanyan, V.}, and {Zargaryan, D.}
\newblock Fermi-lat observation of nonblazar agns.
\newblock \emph{A\&A}, 614:\penalty0 A6, 2018.
\newblock \doi{10.1051/0004-6361/201732304}.
\newblock URL \url{https://doi.org/10.1051/0004-6361/201732304}.

\bibitem[Batista et~al.(2016)Batista, Dundovic, Erdmann, Kampert, Kuempel,
  M{\"u}ller, Sigl, van Vliet, Walz, and Winchen]{batista2016crpropa}
Rafael~Alves Batista, Andrej Dundovic, Martin Erdmann, Karl-Heinz Kampert,
  Daniel Kuempel, Gero M{\"u}ller, Guenter Sigl, Arjen van Vliet, David Walz,
  and Tobias Winchen.
\newblock {CRP}ropa 3 -- a public astrophysical simulation framework for
  propagating extraterrestrial ultra-high energy particles.
\newblock \emph{Journal of Cosmology and Astroparticle Physics}, 2016\penalty0
  (05):\penalty0 038, 2016.

\bibitem[Gilmore et~al.(2012)Gilmore, Somerville, Primack, and
  Dom{\'\i}nguez]{gilmore2012semi}
Rudy~C Gilmore, Rachel~S Somerville, Joel~R Primack, and Alberto
  Dom{\'\i}nguez.
\newblock Semi-analytic modelling of the extragalactic background light and
  consequences for extragalactic gamma-ray spectra.
\newblock \emph{Monthly Notices of the Royal Astronomical Society},
  422\penalty0 (4):\penalty0 3189--3207, 2012.

\bibitem[Bretz et~al.(2014)Bretz, Erdmann, Schiffer, Walz, and
  Winchen]{BRETZ2014110}
Hans-Peter Bretz, Martin Erdmann, Peter Schiffer, David Walz, and Tobias
  Winchen.
\newblock Parsec: A parametrized simulation engine for ultra-high energy cosmic
  ray protons.
\newblock \emph{Astroparticle Physics}, 54:\penalty0 110--117, 2014.
\newblock ISSN 0927-6505.
\newblock \doi{https://doi.org/10.1016/j.astropartphys.2013.12.002}.
\newblock URL
  \url{https://www.sciencedirect.com/science/article/pii/S0927650513001874}.

\bibitem[Lamastra et~al.(2019)Lamastra, Tavecchio, Romano, Landoni, and
  Vercellone]{LAMASTRA201916}
Alessandra Lamastra, Fabrizio Tavecchio, Patrizia Romano, Marco Landoni, and
  Stefano Vercellone.
\newblock Unveiling the origin of the gamma-ray emission in ngc 1068 with the
  cherenkov telescope array.
\newblock \emph{Astroparticle Physics}, 112:\penalty0 16--23, 2019.
\newblock ISSN 0927-6505.
\newblock \doi{https://doi.org/10.1016/j.astropartphys.2019.04.003}.
\newblock URL
  \url{https://www.sciencedirect.com/science/article/pii/S0927650519300362}.

\bibitem[Aublin and Parizot(2005)]{aublin2005generalised}
J~Aublin and E~Parizot.
\newblock Generalised 3d-reconstruction method of a dipole anisotropy in
  cosmic-ray distributions.
\newblock \emph{Astronomy \& Astrophysics}, 441\penalty0 (1):\penalty0
  407--415, 2005.

\bibitem[Sigl et~al.(2004)Sigl, Miniati, and Ensslin]{SIGL2004224}
G\"unter Sigl, Francesco Miniati, and Torsten~A. Ensslin.
\newblock Cosmic magnetic fields and their influence on ultra-high energy
  cosmic ray propagation.
\newblock \emph{Nuclear Physics B - Proceedings Supplements}, 136:\penalty0
  224--233, 2004.
\newblock ISSN 0920-5632.
\newblock \doi{https://doi.org/10.1016/j.nuclphysbps.2004.10.043}.
\newblock URL
  \url{https://www.sciencedirect.com/science/article/pii/S0920563204004608}.
\newblock CRIS 2004 Proceedings of the Cosmic Ray International Seminars: GZK
  and Surroundings.

\bibitem[Tanco(2001)]{Tanco2001}
Gustavo~Medina Tanco.
\newblock \emph{Cosmic Magnetic Fields from the Perspective of
  Ultra-High-Energy Cosmic Rays Propagation}, pages 155--180.
\newblock Springer Berlin Heidelberg, Berlin, Heidelberg, 2001.
\newblock ISBN 978-3-540-45615-5.
\newblock \doi{10.1007/3-540-45615-5_7}.
\newblock URL \url{https://doi.org/10.1007/3-540-45615-5_7}.

\bibitem[Tanco(1998)]{Medina_Tanco_1998}
Gustavo A.~Medina Tanco.
\newblock The effect of highly structured cosmic magnetic fields on
  ultra{\textendash}high-energy cosmic-ray propagation.
\newblock \emph{The Astrophysical Journal}, 505\penalty0 (2):\penalty0
  L79--L82, oct 1998.
\newblock \doi{10.1086/311615}.
\newblock URL \url{https://doi.org/10.1086/311615}.

\bibitem[{Lee} et~al.(1995){Lee}, {Olinto}, and {Sigl}]{1995ApJ...455L..21L}
Sangjin {Lee}, Angela~V. {Olinto}, and Guenter {Sigl}.
\newblock {Extragalactic Magnetic Field and the Highest Energy Cosmic Rays}.
\newblock \emph{The Astrophysical Journal Letter}, 455:\penalty0 L21, December
  1995.
\newblock \doi{10.1086/309812}.

\bibitem[Aab et~al.(2017)Aab, Abreu, Aglietta, Al~Samarai, Albuquerque,
  Allekotte, Almela, Alvarez~Castillo, Alvarez-Mu\~niz, Anastasi, Anchordoqui,
  Andrada, Andringa, Aramo, Arqueros, Arsene, Asorey, Assis, Aublin, Avila,
  Badescu, Balaceanu, Barbato, Barreira~Luz, Beatty, Becker, Bellido, Berat,
  Bertaina, Bertou, Biermann, Biteau, Blaess, Blanco, Blazek, Bleve,
  Boh\'a\v{c}ov\'a, Boncioli, Bonifazi, Borodai, Botti, Brack, Brancus, Bretz,
  Bridgeman, Briechle, Buchholz, Bueno, Buitink, Buscemi, Caballero-Mora,
  Caccianiga, Cancio, Canfora, Caramete, Caruso, Castellina, Catalani, Cataldi,
  Cazon, Chavez, Chinellato, Chudoba, Clay, Cobos, Colalillo, Coleman, Collica,
  Coluccia, Concei\c{c}\~ao, Consolati, Contreras, Cooper, Coutu, Covault,
  Cronin, D'Amico, Daniel, Dasso, Daumiller, Dawson, de~Almeida, de~Jong,
  De~Mauro, de~Mello~Neto, De~Mitri, de~Oliveira, de~Souza, Debatin, Deligny,
  D\'{\i}az~Castro, Diogo, Dobrigkeit, D'Olivo, Dorosti, dos Anjos, Dova,
  Dundovic, Ebr, Engel, Erdmann, Erfani, Escobar, Espadanal, Etchegoyen,
  Falcke, Farmer, Farrar, Fauth, Fazzini, Fenu, Fick, Figueira,
  Filip\v{c}i\v{c}, Fratu, Freire, Fujii, Fuster, Gaior, Garc\'{\i}a,
  Garcia-Pinto, Gat\'e, Gemmeke, Gherghel-Lascu, Ghia, Giaccari, Giammarchi,
  Giller, G\l{}as, Glaser, Golup, G\'omez~Berisso, G\'omez~Vitale, Gonz\'alez,
  Gorgi, Gorham, Grillo, Grubb, Guarino, Guedes, Halliday, Hampel, Hansen,
  Harari, Harrison, Harton, Haungs, Hebbeker, Heck, Heimann, Herve, Hill,
  Hojvat, Holt, Homola, H\"orandel, Horvath, Hrabovsk\'y, Huege, Hulsman,
  Insolia, Isar, Jandt, Johnsen, Josebachuili, Jurysek, K\"a\"ap\"a, Kambeitz,
  Kampert, Keilhauer, Kemmerich, Kemp, Kemp, Kieckhafer, Klages, Kleifges,
  Kleinfeller, Krause, Krohm, Kuempel, Kukec~Mezek, Kunka, Kuotb~Awad, Lago,
  LaHurd, Lang, Lauscher, Legumina, Leigui~de Oliveira, Letessier-Selvon,
  Lhenry-Yvon, Link, Lo~Presti, Lopes, L\'opez, L\'opez~Casado, Lorek, Luce,
  Lucero, Malacari, Mallamaci, Mandat, Mantsch, Mariazzi, Mari\c{s}, Marsella,
  Martello, Martinez, Mart\'{\i}nez~Bravo, Mas\'{\i}as~Meza, Mathes, Mathys,
  Matthews, Matthews, Matthiae, Mayotte, Mazur, Medina, Medina-Tanco, Melo,
  Menshikov, Merenda, Michal, Micheletti, Middendorf, Miramonti, Mitrica,
  Mockler, Mollerach, Montanet, Morello, Mostaf\'a, M\"uller, M\"uller, Muller,
  M\"uller, Mussa, Naranjo, Nellen, Nguyen, Niculescu-Oglinzanu, Niechciol,
  Niemietz, Niggemann, Nitz, Nosek, Novotny, No\v{z}ka, N\'u\~nez, Ochilo,
  Oikonomou, Olinto, Palatka, Pallotta, Papenbreer, Parente, Parra, Paul, Pech,
  Pedreira, P\k{e}kala, Pelayo, Pe\~na Rodriguez, Pereira, Perlin, Perrone,
  Peters, Petrera, Phuntsok, Piegaia, Pierog, Pimenta, Pirronello, Platino,
  Plum, Porowski, Prado, Privitera, Prouza, Quel, Querchfeld, Quinn,
  Ramos-Pollan, Rautenberg, Ravignani, Ridky, Riehn, Risse, Ristori, Rizi,
  Rodrigues~de Carvalho, Rodriguez~Fernandez, Rodriguez~Rojo, Rogozin,
  Roncoroni, Roth, Roulet, Rovero, Ruehl, Saffi, Saftoiu, Salamida, Salazar,
  Saleh, Salesa~Greus, Salina, S\'anchez, Sanchez-Lucas, Santos, Santos,
  Sarazin, Sarmento, Sarmiento-Cano, Sato, Schauer, Scherini, Schieler, Schimp,
  Schmidt, Scholten, Schov\'anek, Schr\"oder, Schr\"oder, Schulz, Schumacher,
  Sciutto, Segreto, Shadkam, Shellard, Sigl, Silli, Sima, \'{S}mia\l{}kowski,
  \v{S}m\'{\i}da, Smith, Snow, Sommers, Sonntag, Squartini, Stanca, Stani\v{c},
  Stasielak, Stassi, Stolpovskiy, Strafella, Streich, Suarez, Suarez~Dur\'an,
  Sudholz, Suomij\"arvi, Supanitsky, \v{S}up\'{\i}k, Swain, Szadkowski,
  Taboada, Taborda, Theodoro, Timmermans, Todero~Peixoto, Tomankova, Tom\'e,
  Torralba~Elipe, Travnicek, Trini, Ulrich, Unger, Urban, Vald\'es~Galicia,
  Vali\~no, Valore, van Aar, van Bodegom, van~den Berg, van Vliet, Varela,
  Vargas~C\'ardenas, Varner, V\'azquez, Veberi\v{c}, Ventura, Vergara~Quispe,
  Verzi, Vicha, Villase\~nor, Vorobiov, Wahlberg, Wainberg, Walz, Watson,
  Weber, Weindl, Wiencke, Wilczy\'{n}ski, Wileman, Wirtz, Wittkowski,
  Wundheiler, Yang, Yushkov, Zas, Zavrtanik, Zavrtanik, Zepeda, Zimmermann,
  Ziolkowski, Zong, and Zuccarello]{PhysRevD.96.122003}
A.~Aab, P.~Abreu, M.~Aglietta, I.~Al~Samarai, I.~F.~M. Albuquerque,
  I.~Allekotte, A.~Almela, J.~Alvarez~Castillo, J.~Alvarez-Mu\~niz, G.~A.
  Anastasi, L.~Anchordoqui, B.~Andrada, S.~Andringa, C.~Aramo, F.~Arqueros,
  N.~Arsene, H.~Asorey, P.~Assis, J.~Aublin, G.~Avila, A.~M. Badescu,
  A.~Balaceanu, F.~Barbato, R.~J. Barreira~Luz, J.~J. Beatty, K.~H. Becker,
  J.~A. Bellido, C.~Berat, M.~E. Bertaina, X.~Bertou, P.~L. Biermann,
  J.~Biteau, S.~G. Blaess, A.~Blanco, J.~Blazek, C.~Bleve, M.~Boh\'a\v{c}ov\'a,
  D.~Boncioli, C.~Bonifazi, N.~Borodai, A.~M. Botti, J.~Brack, I.~Brancus,
  T.~Bretz, A.~Bridgeman, F.~L. Briechle, P.~Buchholz, A.~Bueno, S.~Buitink,
  M.~Buscemi, K.~S. Caballero-Mora, L.~Caccianiga, A.~Cancio, F.~Canfora,
  L.~Caramete, R.~Caruso, A.~Castellina, F.~Catalani, G.~Cataldi, L.~Cazon,
  A.~G. Chavez, J.~A. Chinellato, J.~Chudoba, R.~W. Clay, A.~Cobos,
  R.~Colalillo, A.~Coleman, L.~Collica, M.~R. Coluccia, R.~Concei\c{c}\~ao,
  G.~Consolati, F.~Contreras, M.~J. Cooper, S.~Coutu, C.~E. Covault, J.~Cronin,
  S.~D'Amico, B.~Daniel, S.~Dasso, K.~Daumiller, B.~R. Dawson, R.~M.
  de~Almeida, S.~J. de~Jong, G.~De~Mauro, J.~R.~T. de~Mello~Neto, I.~De~Mitri,
  J.~de~Oliveira, V.~de~Souza, J.~Debatin, O.~Deligny, M.~L. D\'{\i}az~Castro,
  F.~Diogo, C.~Dobrigkeit, J.~C. D'Olivo, Q.~Dorosti, R.~C. dos Anjos, M.~T.
  Dova, A.~Dundovic, J.~Ebr, R.~Engel, M.~Erdmann, M.~Erfani, C.~O. Escobar,
  J.~Espadanal, A.~Etchegoyen, H.~Falcke, J.~Farmer, G.~Farrar, A.~C. Fauth,
  N.~Fazzini, F.~Fenu, B.~Fick, J.~M. Figueira, A.~Filip\v{c}i\v{c}, O.~Fratu,
  M.~M. Freire, T.~Fujii, A.~Fuster, R.~Gaior, B.~Garc\'{\i}a, D.~Garcia-Pinto,
  F.~Gat\'e, H.~Gemmeke, A.~Gherghel-Lascu, P.~L. Ghia, U.~Giaccari,
  M.~Giammarchi, M.~Giller, D.~G\l{}as, C.~Glaser, G.~Golup,
  M.~G\'omez~Berisso, P.~F. G\'omez~Vitale, N.~Gonz\'alez, A.~Gorgi, P.~Gorham,
  A.~F. Grillo, T.~D. Grubb, F.~Guarino, G.~P. Guedes, R.~Halliday, M.~R.
  Hampel, P.~Hansen, D.~Harari, T.~A. Harrison, J.~L. Harton, A.~Haungs,
  T.~Hebbeker, D.~Heck, P.~Heimann, A.~E. Herve, G.~C. Hill, C.~Hojvat,
  E.~Holt, P.~Homola, J.~R. H\"orandel, P.~Horvath, M.~Hrabovsk\'y, T.~Huege,
  J.~Hulsman, A.~Insolia, P.~G. Isar, I.~Jandt, J.~A. Johnsen, M.~Josebachuili,
  J.~Jurysek, A.~K\"a\"ap\"a, O.~Kambeitz, K.~H. Kampert, B.~Keilhauer,
  N.~Kemmerich, E.~Kemp, J.~Kemp, R.~M. Kieckhafer, H.~O. Klages, M.~Kleifges,
  J.~Kleinfeller, R.~Krause, N.~Krohm, D.~Kuempel, G.~Kukec~Mezek, N.~Kunka,
  A.~Kuotb~Awad, B.~L. Lago, D.~LaHurd, R.~G. Lang, M.~Lauscher, R.~Legumina,
  M.~A. Leigui~de Oliveira, A.~Letessier-Selvon, I.~Lhenry-Yvon, K.~Link,
  D.~Lo~Presti, L.~Lopes, R.~L\'opez, A.~L\'opez~Casado, R.~Lorek, Q.~Luce,
  A.~Lucero, M.~Malacari, M.~Mallamaci, D.~Mandat, P.~Mantsch, A.~G. Mariazzi,
  I.~C. Mari\c{s}, G.~Marsella, D.~Martello, H.~Martinez,
  O.~Mart\'{\i}nez~Bravo, J.~J. Mas\'{\i}as~Meza, H.~J. Mathes, S.~Mathys,
  J.~Matthews, J.~A.~J. Matthews, G.~Matthiae, E.~Mayotte, P.~O. Mazur,
  C.~Medina, G.~Medina-Tanco, D.~Melo, A.~Menshikov, K.-D. Merenda, S.~Michal,
  M.~I. Micheletti, L.~Middendorf, L.~Miramonti, B.~Mitrica, D.~Mockler,
  S.~Mollerach, F.~Montanet, C.~Morello, M.~Mostaf\'a, A.~L. M\"uller,
  G.~M\"uller, M.~A. Muller, S.~M\"uller, R.~Mussa, I.~Naranjo, L.~Nellen,
  P.~H. Nguyen, M.~Niculescu-Oglinzanu, M.~Niechciol, L.~Niemietz,
  T.~Niggemann, D.~Nitz, D.~Nosek, V.~Novotny, L.~No\v{z}ka, L.~A. N\'u\~nez,
  L.~Ochilo, F.~Oikonomou, A.~Olinto, M.~Palatka, J.~Pallotta, P.~Papenbreer,
  G.~Parente, A.~Parra, T.~Paul, M.~Pech, F.~Pedreira, J.~P\k{e}kala,
  R.~Pelayo, J.~Pe\~na Rodriguez, L.~A.~S. Pereira, M.~Perlin, L.~Perrone,
  C.~Peters, S.~Petrera, J.~Phuntsok, R.~Piegaia, T.~Pierog, M.~Pimenta,
  V.~Pirronello, M.~Platino, M.~Plum, C.~Porowski, R.~R. Prado, P.~Privitera,
  M.~Prouza, E.~J. Quel, S.~Querchfeld, S.~Quinn, R.~Ramos-Pollan,
  J.~Rautenberg, D.~Ravignani, J.~Ridky, F.~Riehn, M.~Risse, P.~Ristori,
  V.~Rizi, W.~Rodrigues~de Carvalho, G.~Rodriguez~Fernandez, J.~Rodriguez~Rojo,
  D.~Rogozin, M.~J. Roncoroni, M.~Roth, E.~Roulet, A.~C. Rovero, P.~Ruehl,
  S.~J. Saffi, A.~Saftoiu, F.~Salamida, H.~Salazar, A.~Saleh, F.~Salesa~Greus,
  G.~Salina, F.~S\'anchez, P.~Sanchez-Lucas, E.~M. Santos, E.~Santos,
  F.~Sarazin, R.~Sarmento, C.~Sarmiento-Cano, R.~Sato, M.~Schauer, V.~Scherini,
  H.~Schieler, M.~Schimp, D.~Schmidt, O.~Scholten, P.~Schov\'anek, F.~G.
  Schr\"oder, S.~Schr\"oder, A.~Schulz, J.~Schumacher, S.~J. Sciutto,
  A.~Segreto, A.~Shadkam, R.~C. Shellard, G.~Sigl, G.~Silli, O.~Sima,
  A.~\'{S}mia\l{}kowski, R.~\v{S}m\'{\i}da, B.~Smith, G.~R. Snow, P.~Sommers,
  S.~Sonntag, R.~Squartini, D.~Stanca, S.~Stani\v{c}, J.~Stasielak, P.~Stassi,
  M.~Stolpovskiy, F.~Strafella, A.~Streich, F.~Suarez, M.~Suarez~Dur\'an,
  T.~Sudholz, T.~Suomij\"arvi, A.~D. Supanitsky, J.~\v{S}up\'{\i}k, J.~Swain,
  Z.~Szadkowski, A.~Taboada, O.~A. Taborda, V.~M. Theodoro, C.~Timmermans,
  C.~J. Todero~Peixoto, L.~Tomankova, B.~Tom\'e, G.~Torralba~Elipe,
  P.~Travnicek, M.~Trini, R.~Ulrich, M.~Unger, M.~Urban, J.~F.
  Vald\'es~Galicia, I.~Vali\~no, L.~Valore, G.~van Aar, P.~van Bodegom, A.~M.
  van~den Berg, A.~van Vliet, E.~Varela, B.~Vargas~C\'ardenas, G.~Varner, R.~A.
  V\'azquez, D.~Veberi\v{c}, C.~Ventura, I.~D. Vergara~Quispe, V.~Verzi,
  J.~Vicha, L.~Villase\~nor, S.~Vorobiov, H.~Wahlberg, O.~Wainberg, D.~Walz,
  A.~A. Watson, M.~Weber, A.~Weindl, L.~Wiencke, H.~Wilczy\'{n}ski, C.~Wileman,
  M.~Wirtz, D.~Wittkowski, B.~Wundheiler, L.~Yang, A.~Yushkov, E.~Zas,
  D.~Zavrtanik, M.~Zavrtanik, A.~Zepeda, B.~Zimmermann, M.~Ziolkowski, Z.~Zong,
  and F.~Zuccarello.
\newblock Inferences on mass composition and tests of hadronic interactions
  from 0.3 to 100 eev using the water-cherenkov detectors of the pierre auger
  observatory.
\newblock \emph{Phys. Rev. D}, 96:\penalty0 122003, Dec 2017.
\newblock \doi{10.1103/PhysRevD.96.122003}.
\newblock URL \url{https://link.aps.org/doi/10.1103/PhysRevD.96.122003}.

\bibitem[Hillas(2006)]{bib:hillas}
A.~M. Hillas.
\newblock Cosmic rays: Recent progress and some current questions, 2006.
\newblock URL \url{https://arxiv.org/abs/astro-ph/0607109}.

\bibitem[Peixoto et~al.(2015)Peixoto, de~Souza, and Biermann]{Peixoto_2015}
C.J.~Todero Peixoto, Vitor de~Souza, and Peter~L. Biermann.
\newblock Cosmic rays: the spectrum and chemical composition from $10^{10}$ to
  $10^{20}${eV}.
\newblock \emph{Journal of Cosmology and Astroparticle Physics}, 2015\penalty0
  (07):\penalty0 042--042, jul 2015.
\newblock \doi{10.1088/1475-7516/2015/07/042}.
\newblock URL \url{https://doi.org/10.1088/1475-7516/2015/07/042}.

\bibitem[Paczynski(1998)]{bib:paczynski}
Bohdan Paczynski.
\newblock Gamma-ray bursts as hypernovae.
\newblock \emph{Fourth Huntsville gamma-ray burst symposium. AIP Conference
  Proceedings}, 428, 1998.

\bibitem[Wang et~al.(2008)Wang, Razzaque, and M\'esz\'aros]{bib:wang}
Xiang-Yu Wang, Soebur Razzaque, and Peter M\'esz\'aros.
\newblock On the origin and survival of ultra-high-energy cosmic-ray nuclei in
  gamma-ray bursts and hypernovae.
\newblock \emph{The Astrophysical Journal}, 677:\penalty0 432, 2008.

\bibitem[Blasi et~al.(2000)Blasi, Epstein, and Olinto]{Blasi_2000}
P.~Blasi, R.~I. Epstein, and A.~V. Olinto.
\newblock Ultra{\textendash}high-energy cosmic rays from young neutron star
  winds.
\newblock \emph{The Astrophysical Journal}, 533\penalty0 (2):\penalty0
  L123--L126, apr 2000.
\newblock \doi{10.1086/312626}.
\newblock URL \url{https://doi.org/10.1086/312626}.

\bibitem[Aloisio et~al.(2014)Aloisio, Berezinsky, and
  Blasi]{1475-7516-2014-10-020}
R.~Aloisio, V.~Berezinsky, and P.~Blasi.
\newblock Ultra high energy cosmic rays: implications of auger data for source
  spectra and chemical composition.
\newblock \emph{Journal of Cosmology and Astroparticle Physics}, 2014\penalty0
  (10):\penalty0 020, 2014.
\newblock URL \url{http://stacks.iop.org/1475-7516/2014/i=10/a=020}.

\end{thebibliography}

%-------------------------------------
\newpage
\appendix
\section{Effect of the GMF over the hotspots}
\label{sec:appendix}

\begin{figure}
  \centering
  \includegraphics[width=1.0\columnwidth]{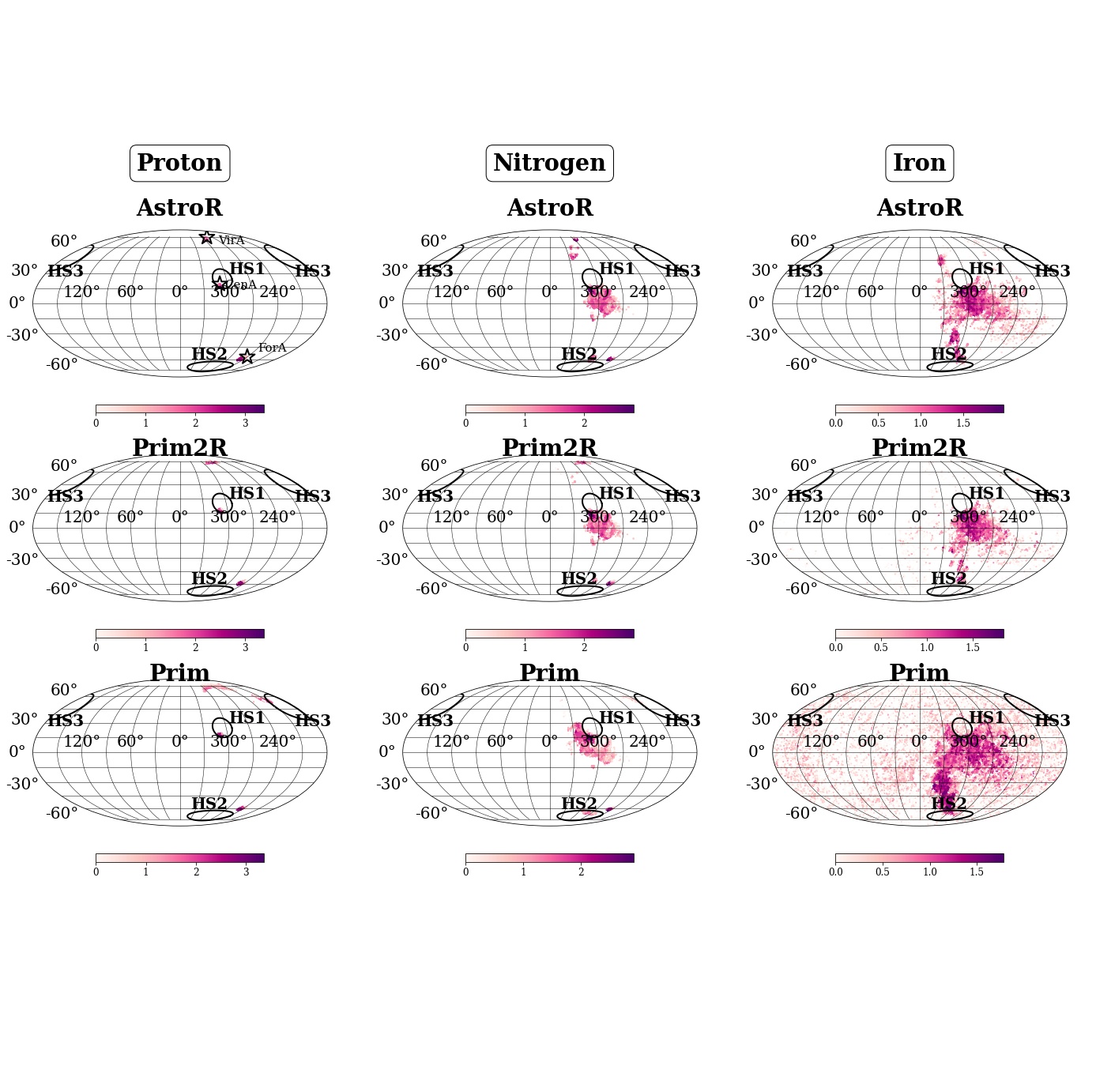}
  \caption{Arrival directions map (galactic coordinates) in a Mollweide projection for events with energies above 60~EeV arriving at Earth injected by the AGNs (Cen~A, Vir~A, and For~A). The maps correspond to the 1:1 UHECR luminosity. The colorscale indicates the number of events on logarithm scale. In each line, an EGMF model is presented: AstroR, Prim2R, and Prim. In each column, the injected composition is shown: proton, nitrogen, and iron. The position of the sources (AGNs) is shown only in the first plot for sake of clarity. The hotspot regions measured by the Pierre Auger (HS1 and HS2) and the Telescope Array (HS3) are also shown.}
  \label{fig:arrival:60EeV:AGN}
\end{figure}

\begin{figure}
  \centering
  \includegraphics[width=1.0\columnwidth]{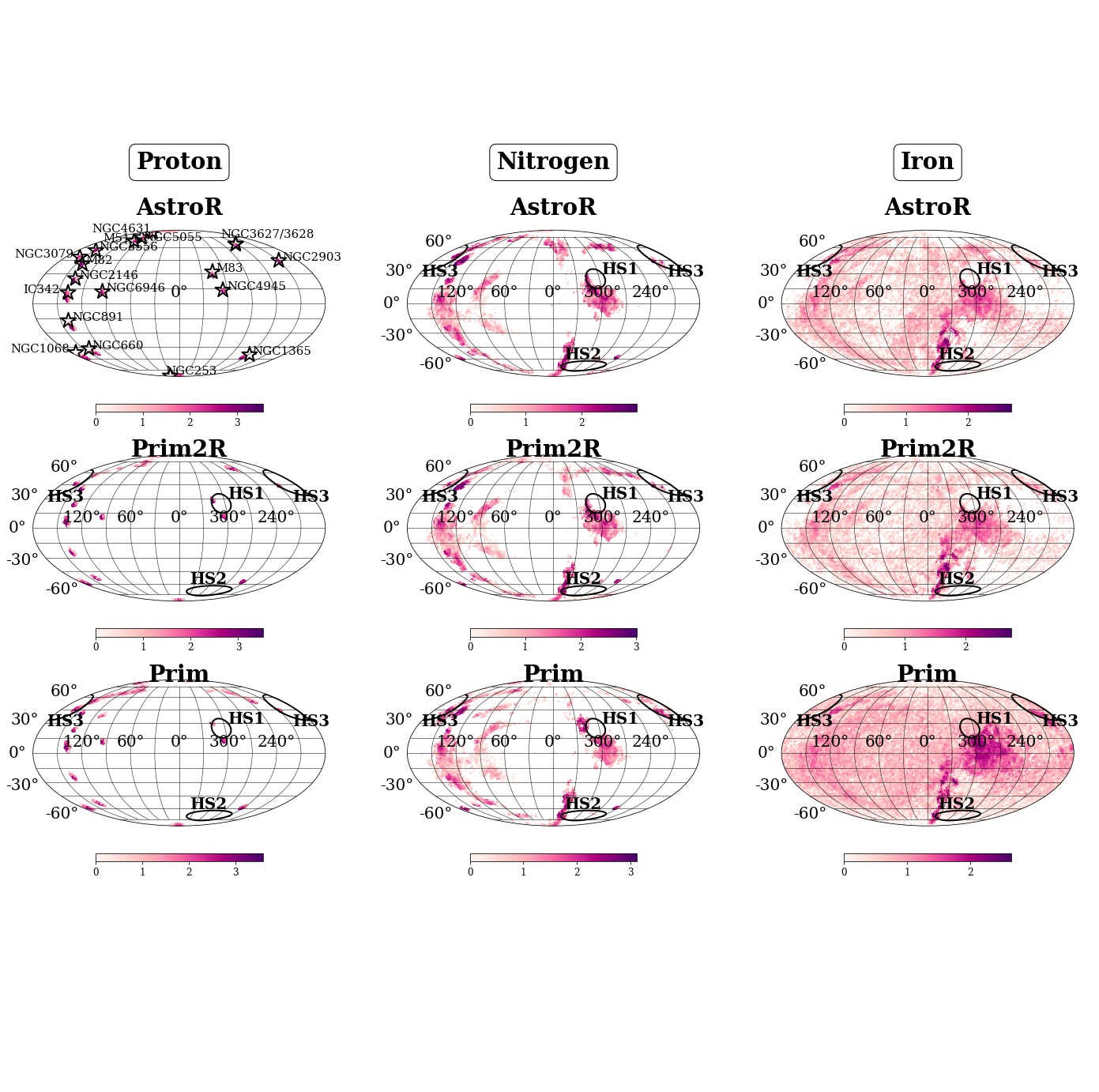}
  \caption{Arrival directions map in a Mollweide projection for events with energies above 60~EeV arriving at Earth injected by the SBGs in Table~\ref{tab:sources}. The maps correspond to the 1:1 UHECR luminosity. The elements of figure are the same of figure~\ref{fig:arrival:60EeV:AGN}.}
  \label{fig:arrival:60EeV:SBG}
\end{figure}

\begin{figure}
  \centering
  \includegraphics[width=1.0\columnwidth]{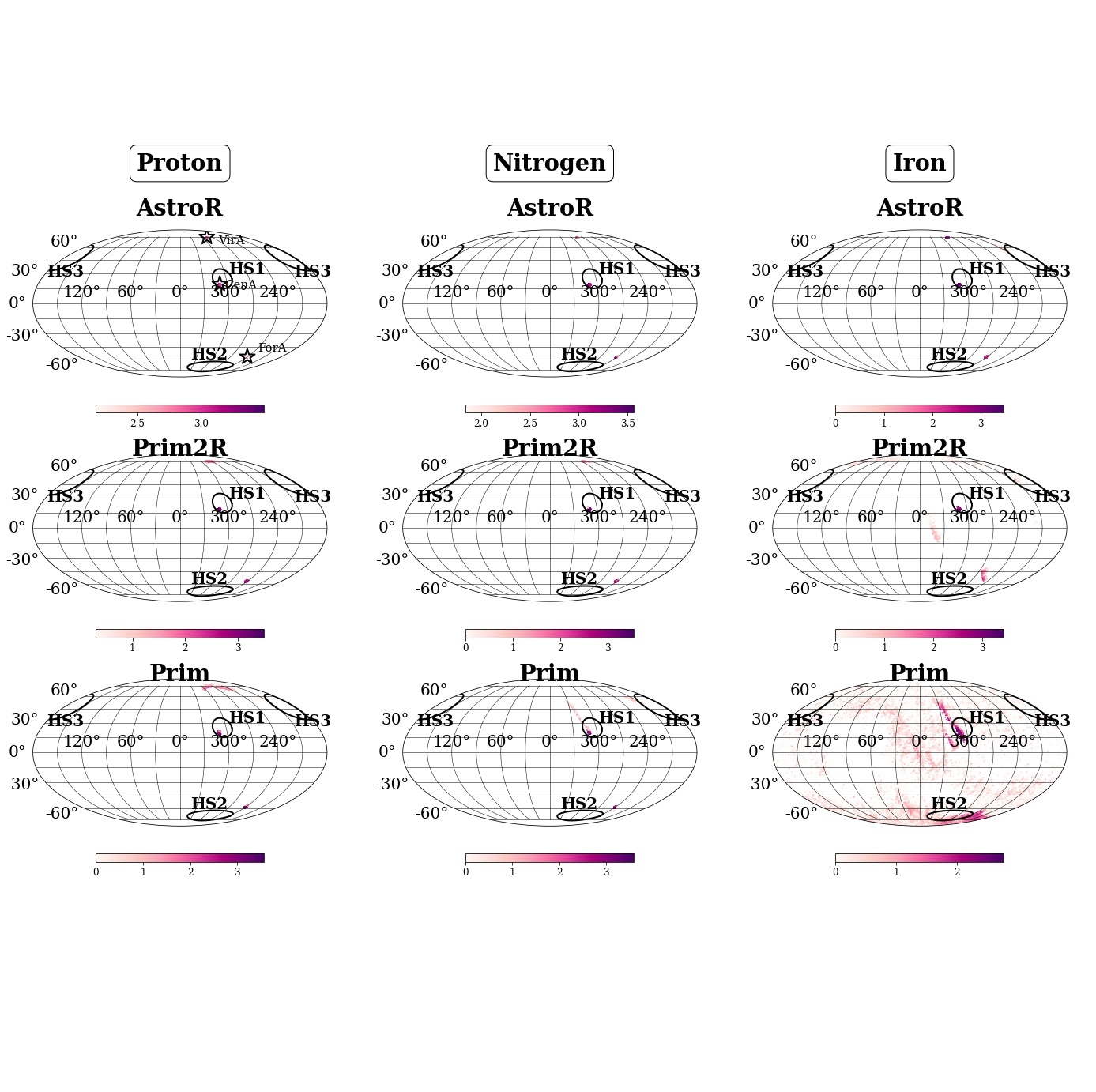}
  \caption{Arrival directions map for events with energies above 60~EeV for the AGNs in a Mollweide projection. The GMF effect is not included. The elements of figure are the same of figure \ref{fig:arrival:60EeV:AGN}.}
  \label{fig:arrival:60EeV:AGN:noGMF}
\end{figure}

\begin{figure}
  \centering
  \includegraphics[width=1.0\columnwidth]{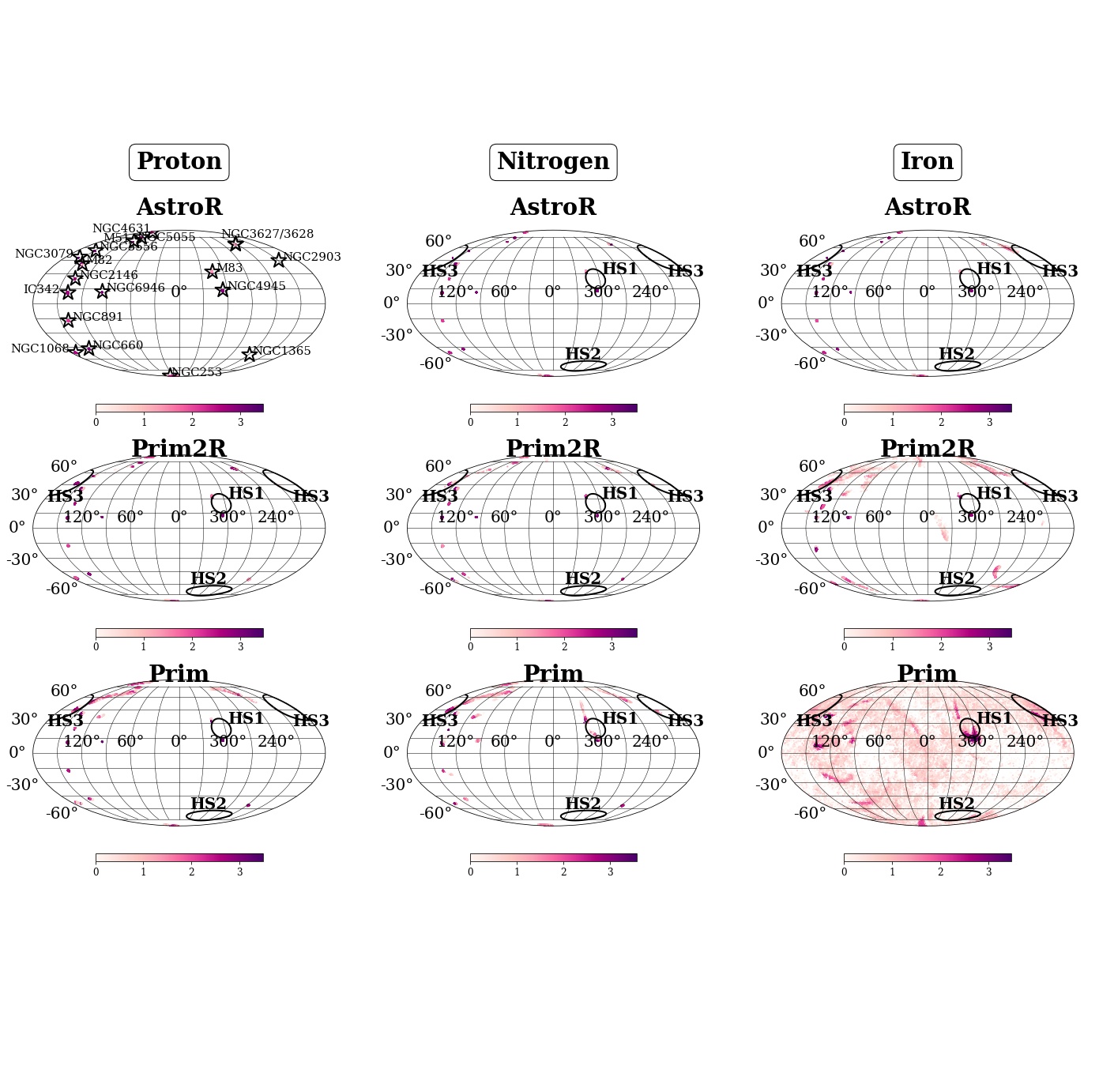}
  \caption{Arrival directions map for events with energies above 60~EeV for the SBGs in a Mollweide projection. The GMF effect is not included. The elements of figure are the same of figure \ref{fig:arrival:60EeV:AGN}.}
  \label{fig:arrival:60EeV:SBG:noGMF}
\end{figure}

To ilustrate the effect of the GMF, we compare the arrival direction of UHECR with energies above $60$~EeV in the presence and in the absence of the GMF. Figure \ref{fig:arrival:60EeV:AGN:noGMF} and \ref{fig:arrival:60EeV:SBG:noGMF} shows the arrival directions map for events with energies above 60~EeV when local AGNs and SBGs were considered the sources of UHECR, respectively. The GMF effect is not included in these sky maps. These figures are analogues to the figures \ref{fig:arrival:60EeV:AGN} and \ref{fig:arrival:60EeV:SBG}, in which the effect of the GMF is included. Comparing the figures, the effect of the GMF over the regions with excess of events from one particular source can be observed. In all the cases, it is possible to note that the GMF is responsible for blur the signal, since the arrival direction in the absence of the GMF is much more point-like.

Comparing figures \ref{fig:arrival:60EeV:AGN:noGMF} and \ref{fig:arrival:60EeV:AGN}, we can obtain the effects causing in the AGN cases. In the absence of the GMF, the events coming from Cen~A majority are inside the HS1 region. The GMF acts attracting the events of Cen~A to lower latitudes, removing then from the HS1 region. The GMF tends to attract the Vir~A signal to lower latitudes in the cases in which the EGMF causes a small deflection, and to increase the deflection in direction to the HS3 region in the case of a large shift due to the EGMF. Over the events coming from For~A, the GMF attract then to higher latitude, contributing to the population of the HS2.

Comparing figures \ref{fig:arrival:60EeV:SBG:noGMF} and \ref{fig:arrival:60EeV:SBG}, we can obtain the effects causing in the SBG cases. The most pronuncied effect is to blur the events of the majority of sources. The GMF act over the events of M83 attracting the light fraction inside the HS1 and blurring the heavier fraction. In addition, the GMF relocate the contribution of NGC~4945, pulling events from that source from the direction of HS1. The GMF is also the responsible for attract the events from NGC~253 to the direction of the HS2, forming a large strip coming from lower latitudes.

%-------------------------------------

%===========================

\begin{table}[htb]
    \caption{Properties of the sources considered: source name, distance, the radio luminosity at 1.4 GHz ($L_{radio}$) and the $\gamma$-ray luminosity between 0.1 and 100 GeV ($L_{\gamma}$).}
\begin{center}
\begin{tabular}{ |c|c|c|c| }
 \hline
 Source & Distance (Mpc) & $L_{radio}$ ($10^{38}$ erg s$^{-1}$) & $L_{\gamma}$ ($10^{40}$ erg s$^{-1}$)\\
 \hline
 NGC 253 & 2.7 & 1.0 & 0.8\\
 M82 & 3.6 & 1.3 & 1.7\\
 NGC 4945 & 4.0 & 1.0 & 1.4\\
 M83 & 4.0 & 0.4 & 1.0\\
 IC 342 & 4.0 & 0.5 & 0.4\\
 NGC 6946 & 5.9 & 0.7 & 0.5\\
 NGC 2903 & 6.6 & 0.7 & 0.9\\
 NGC 5055 & 7.8 & 0.6 & 1.1\\
 NGC 3628 & 8.1 & 1.0 & 1.8\\
 NGC 3627 & 8.1 & 0.7 & 2.0\\
 NGC 4631 & 8.7 & 1.1 & 1.0\\
 M51 & 10.3 & 1.2 & 2.9\\
 NGC 891 & 11 & 0.9 & 4.4\\
 NGC 3556 & 11.4 & 0.8 & 2.6\\
 NGC 660 & 15 & 0.9 & 5.8\\
 NGC 2146 & 16.3 & 4.1 & 15.4\\
 NGC 3079 & 17.4 & 5.0 & 5.7\\
 NGC 1068 & 17.9 & 17.8 & 17.7\\
 NGC 1365 & 22.3 & 3.1 & 10.7\\
 \hline
 \hline
 Centaurus A & 3.8 & 260 & 11.2\\
 Virgo A & 18.4 & 760 & 78.9\\
 Fornax A & 20.9 & 830 & 27.8\\
 \hline
\end{tabular}
\label{tab:sources}
\end{center}
\end{table}

\newpage
%======================================
%Hotspots generated by AGNs and/or SBGs}

%END: Hotspots generated by AGNs and/or SBGs}
%----------------------------------------------------------------
%-----------------------------------------------------------
%Dipole generated by AGNs and/or SBGs

%\begin{figure}
%  \centering
%  \includegraphics[width=1.0\columnwidth]{dip_8EeV_SBG-AGN.pdf}
%  \caption{Dipole direction generated by events with energy above 8~EeV if SBGs and AGNs are considered sources. Each column presents different UHECR luminosities proxies: 1:1, Radio, and Gamma. Each line are showns different EGMF models: AstroR, Prim2R and Prim. The nuclei emitted by the source is mapped by colors: proton (p, red), helium (He, orange), nitrogen (N, green), silicon (Si, cyan), and iron (Fe, magenta). The brown square presents the dipole direction measured by the Pierre Auger Observatory~\citep{deAlmeida:20212Z}. The map is in galactic coordinates.}
%  \label{fig:dip_8EeV_SBG-AGN}
%\end{figure}

%\begin{figure}
%  \centering
%  \includegraphics[width=1.0\columnwidth]{dip_8EeV_SBG.pdf}
%  \caption{Dipole direction generated by events with energy above 8~EeV if SBGs are considered sources. The elements in the figure are the same of figure~\ref{fig:dip_8EeV_SBG-AGN}.}
%  \label{fig:dip_8EeV_SBG}
%\end{figure}

%\begin{figure}
%  \centering
%  \includegraphics[width=1.0\columnwidth]{dip_32EeV_SBG.pdf}
%  \caption{Dipole direction generated by events with energy above 32~EeV if SBGs are considered sources. The elements in the figure are the same of figure~\ref{fig:dip_8EeV_SBG-AGN}.}
%  \label{fig:dip_32EeV_SBG}
%\end{figure}
%\begin{figure}
%  \centering
%  \includegraphics[width=1.0\columnwidth]{dip_8EeV_AGN.pdf}
%  \caption{Dipole direction generated by events with energy above 8~EeV if AGNs are considered sources. The elements in the figure are the same of figure~\ref{fig:dip_8EeV_SBG-AGN}.}
%  \label{fig:dip_8EeV_AGN}
%\end{figure}

%\begin{figure}
%  \centering
%  \includegraphics[width=1.0\columnwidth]{dip_32EeV_AGN.pdf}
%  \caption{Dipole direction generated by events with energy above 32~EeV if AGNs are considered sources. The elements in the figure are the same of figure~\ref{fig:dip_8EeV_SBG-AGN}.}
%  \label{fig:dip_32EeV_AGN}
%\end{figure}
%-------------------------------------------
%Searching

%-----Appendix Figures------

\end{document}